\documentclass[a4paper,11pt]{article}
\pdfoutput=1 

\usepackage{jheppub}
\usepackage{color}

\usepackage{tensor}
\usepackage{subcaption}
\usepackage[section]{placeins}

\newcommand{\nn}{\nonumber}

\newcommand{\mywidthb}{0.32}
\newcommand{\mywidthc}{0.22}

\def\myf{f}
\def\mydf{\dot f}
\def\myddf{\ddot f}

\title{\boldmath 
The Large Dimension Limit of a Small Black Hole Instability in Anti-de Sitter Space
}

\author[]{Christopher P. Herzog and}
\author[]{Youngshin Kim}

\preprint{YITP-17-49}

\affiliation[]{C.~N.\ Yang Institute for Theoretical Physics, Department of Physics and Astronomy,\\
Stony Brook University, Stony Brook, NY 11794--3840, USA}

\abstract{   
We study the dynamics of a
black hole in an asymptotically $AdS_d \times S^d$ space-time in the limit of a large number of dimensions, $d \to \infty$. 
Such a black hole is known to become dynamically unstable below a critical radius.  We derive the dispersion relation for the quasinormal
mode that governs this instability in an expansion in $1/d$.  We also provide a full nonlinear analysis of the instability at leading order in $1/d$. 
We find solutions that resemble the lumpy black spots and black belts previously constructed numerically for small $d$, breaking the $SO(d+1)$ rotational symmetry of the sphere down to $SO(d)$.
We are also able to follow the time evolution of the instability.  Due possibly to limitations in our analysis, our time dependent simulations do not settle down to stationary solutions.
This work has relevance for strongly interacting gauge theories; 
through the AdS/CFT correspondence, the special case $d=5$ corresponds to maximally supersymmetric Yang-Mills theory on a spatial
$S^3$ in the microcanonical ensemble and in a strong coupling and large number of colors limit.
}

\begin{document} 
\maketitle
\flushbottom

\section{Introduction}

The most well-understood example of gauge/gravity duality is the correspondence between type IIB string theory on $AdS_5 \times S^5$ and $\mathcal{N}=4$ super Yang-Mills with $SU(N)$ gauge group \cite{Maldacena:1997re,Witten:1998qj,Gubser:1998bc}. In the large $N$ and large 't Hooft coupling $\lambda$ limit, the bulk theory reduces to classical gravity. Using general relativity as a tool, gauge/gravity duality has led to better understanding of strongly interacting field theories in general and ${\mathcal N} = 4$ super Yang-Mills in particular. 
An early success of the correspondence was the understanding of a deconfinement type phase transition in ${\mathcal N}=4$ Yang-Mills on a sphere as a Hawking-Page phase transition between thermal $AdS$ in global coordinates and an asymptotically $AdS$ geometry containing a black hole 
 \cite{Witten:1998zw}.
While this thermodynamic phase transition is a well established story in the canonical ensemble, 
in the micro-canonical ensemble the situation is much less clear.  
In the dual geometry, there exist small black holes which are both thermodynamically and dynamically unstable.  
In analogy with the Gregory-Laflamme instability for black strings, 
refs. \cite{Banks:1998dd,Peet:1998cr} conjectured the endpoint of the instability to be a black hole with $S^{8}$ topology, localized on the $S^5$,
with a corresponding spontaneous breaking of the $SO(6)$ R-symmetry to SO(5).  
Later numerical analyses \cite{Hubeny:2002xn,Dias:2015pda,Buchel:2015gxa} pinned down the onset of a Gregory-Laflamme like 
instability as a function of horizon radius.
The endpoint of the classical instability has as yet been studied only numerically \cite{Dias:2015pda,Dias:2016eto}.
The goal of this work is to better understand both the onset and ultimate endpoint of the dynamical instability by taking a novel limit -- the number of dimensions to be large.

While increasing the number of dimensions typically makes Einstein's equations more difficult to solve, simplifications emerge if the 
number $d$ is taken large enough while at the same time imposing that the solution maintain a high degree of symmetry.  In a large dimension limit,
the gravitational effects of a black hole horizon die off very quickly with distance.  The radial direction can be ``integrated out'' of Einstein's equations,
leaving a simpler set of hydrodynamic like equations to solve that govern the physics of a membrane-like horizon.
Such an approximation has already proven useful for studying classical black holes \cite{Emparan:2013moa, Emparan:2014aba, Emparan:2013xia, Emparan:2015hwa}. The approximation has been successfully applied to black holes with planar topology in the context of applied holography \cite{Emparan:2013oza,Andrade:2015hpa,Romero-Bermudez:2015bma}, fluid/gravity duality \cite{Emparan:2015rva, Emparan:2015gva, Herzog:2016hob, Rozali:2017bll, Miyamoto:2017ozn}, and the membrane paradigm \cite{Emparan:2016sjk, Bhattacharyya:2015dva, Dandekar:2016jrp,Dandekar:2016fvw}.  Relevant for us, the formalism can also be extended to black holes with spherical topology \cite{Emparan:2014cia, Emparan:2014jca, Bhattacharyya:2015fdk, Bhattacharyya:2016nhn, Bhattacharyya:2017hpj}.

In this paper, we apply the large dimension limit to a study of black holes in an asymptotically $AdS_d \times S^d$ space-time.  As we want the $d=5$ case to be dual to ${\mathcal N}=4$ Yang-Mills, we include an anti-symmetric $d$-form field strength $F_{d}$ in the equations of motion.  
We start with a linearized analysis of the black hole instability, governed by coupled fluctuations of the metric and $d$-form.  
Through a numerical zero mode analysis in $d=5$, ref.\ \cite{Hubeny:2002xn} calculated the onset of the instability 
as a function of black hole radius.
Several years later, 
ref.\ \cite{Buchel:2015gxa} provided a more detailed numerical 
analysis of the associated quasinormal modes, 
providing plots of the dependence of the mode frequency on horizon radius and 
revealing that the frequency 
is purely imaginary.  
Our large $d$ limit allows us to provide analytic power series 
expressions for these quantities.  
We give the threshhold radius to $O(d^{-3})$ and the quasinormal mode dispersion relation at leading order in a large dimension limit.
Generalizing the numerical analyses of refs.\ \cite{Hubeny:2002xn,Buchel:2015gxa} to $d > 5$, we find our series expressions give good
approximations of the numerical results.

Next, we investigate the full non-linear time evolution of these small black holes at leading order in our large $d$ expansion.
Assuming that the endpoint of the black hole instability preserves an $SO(d)$ symmetry, we take an ansatz for the metric and $d$-form that depends
 only on 
time $t$ and radial coordinate $r$ of $AdS_d$ and a polar angle 
$\theta$ on the $S^d$.  
In fact, we consider two slightly different large $d$ expansions.  In the first, we keep $\theta$ as a dynamical variable while in the second, we
rescale the polar angle $u \sim \sqrt{n}(\theta - \pi/2)$ to zoom in on the dynamics at the equator of the $S^d$.
Ultimately, we derive two sets of hydrodynamic-like partial differential equations, one that depends on $(t, \theta)$ and another on $(t, u)$.

We find a large class of exact solutions of the $(t, \theta)$-system.  At a discrete set of horizon radii, these solutions can be static.  The discrete radii correspond to the onset of the initial Gregory-Laflamme like instability and further instabilities at yet smaller radii.   In general, however, the solutions we find are strongly time dependent, with a $e^{e^t}$ type functional form.
The $(t,u)$-system is more complicated.  We can find simple formulae for static solutions which exist not just at discrete values of the horizon radius but over a range of values.  We can also analyze the time dependence of the system numerically.

The solutions we find share qualitative similarities with 
 numerical studies \cite{Dias:2015pda,Dias:2016eto} 
in $d=5$, which use the de Turck method to search for stationary solutions of Einstein's equation.  These studies have shown evidence for several different types of solutions.  There are fully localized spot and belt type solutions on the $S^5$ which have horizon topology $S^8$ and $S^4 \times S^4$ respectively.  There are also ``lumpy'' solutions which share the symmetries of the localized solutions but which maintain the original $S^3 \times S^5$ horizon topology.  
In our large $d$ limit, our solutions are of the ``lumpy'' variety with 
$S^{d-2} \times S^d$ topology.  
While strictly speaking, the topology of all of our solutions remains $S^{d-2} \times S^d$, the solutions can be in some sense 
``exponentially close'' to fully localized solutions with $S^{2d-2}$ and $S^{d-1} \times S^{d-1}$ topology.

In our analysis, energy and entropy are equal at leading order in our large $d$ expansion.  Thus, we are not able to distinguish the lumpy and homogeneous solutions based on thermodynamic considerations alone.  In the future, we hope to distinguish entropy and energy by extending our analysis to subleading orders.  In the meantime, we try to gain some insight into which configurations are preferred by examining the time evolution of perturbations of the homogeneous solutions.  
We find that at long times, there is run-away behavior, with the horizon radius either growing or shrinking uncontrollably at the poles of the $S^d$.  
Conceivably subleading terms in the $1/d$ expansion could stabilize the runaway behavior.  Alternately, a logical possibility remains that the evolution does not settle down to an equilibrium configuration \cite{Yaffe:2017axl}.

The paper is organized as follows.  In section \ref{sec:2}, we study linear fluctuations of the small black holes in a large $d$ limit.  In section \ref{sec:3}, we provide our nonlinear analysis.  Section \ref{sec:summary} contains a brief discussion and conclusion.  Appendices contain various auxiliary technical results.

\section{Quasinormal modes of black holes in a large $d$ limit}
\label{sec:2}
\subsection{Linearized equations of motion}

We begin with the formulation of the small black hole solutions in an arbitrary number of dimensions. In the special case of ten dimensions, this spacetime can be formed as a solution of the type IIB supergravity equations of motion with the self dual 5-form switched on. 
In arbitrary dimension, we write the action as
\begin{equation}
S = \int d^p x d^q y \sqrt{-g} \left(R - \frac{1}{2q!} |F_q|^2 \right).
\end{equation}
The equations of motion take the form 
\begin{align}
R_{MN} &= \frac{1-q}{2(D-2)q!} g_{MN} |F|^2 + \frac{1}{2(q-1)!}F_{MP_2 ... P_q}\tensor{F}{_N^{P_2...P_q}},\\\nn
d*F_q &= 0.
\label{eq:eom}
\end{align}
In compliance with supergravity, we require $F$ to be self-dual, which means that we are limited to $d \equiv p=q$, and the norm $|F|$ vanishes. The self-dual field vanishes for $d$ even, so we assume $d$ is odd for the purposes of deriving the equations of motion. 
(In the final effective equations, $d$ will be a numerical parameter that can take any real value, however.)  
The canonical stationary solution for this setup is the anti-de Sitter-Schwarzschild metric smeared over a sphere
\begin{align}
ds^2 &= - f(r) dt^2 + \frac{1}{f(r)} dr^2 + r^2 d\Omega^2_{d-2} + L^2 d\Omega^2_d,\\
F_d &= \frac{2(d-1)}{L^2} \left( \mbox{vol}({\mbox{AdS}_d})+\mbox{vol}({S^d}) \right) ,
\end{align}
where
\begin{equation}
f(r) = 1 + \frac{r^2}{L^2} - \left( \frac{r_+}{r} \right)^{d-3} \left( 1+\frac{r_+^2}{L^2} \right).
\end{equation}
The line element $d \Omega_d^2$ is constructed from the usual metric on a $d$-dimensional sphere of radius one.

We would like to consider linearized fluctuations over this background. We choose a transverse-traceless gauge, in which the fluctuation of the $d$-form field can be set to zero consistently. For the metric fluctuation $g_{MN} = g_{MN}^{(0)} + h_{MN}$, we consider spherical harmonic perturbations $Y_{\ell_1 \cdots \ell_d} (\Omega)$  on the internal sphere $\Omega_d$. The eigenvalue of the Laplacian will depend on only $\ell_1 
\equiv \ell$.  (See appendix \ref{app:spher} for more details.) The line element can be put in the form 
\begin{align}
\label{eq:fluc}
ds^2 = -f(r)(1 &+ \epsilon e^{-i \omega t} \chi_1(r) Y_\ell(\Omega_d)) dt^2 + \frac{1}{f(r)} (1 + \epsilon e^{-i \omega t} \chi_2(r) Y_\ell (\Omega_d)) dr^2\\\nn
&+ 2 \epsilon e^{-i\omega t} \chi_3(r) Y_\ell (\Omega_d) dt dr + r^2 (1 + \epsilon e^{-i \omega t} \chi_4(r) Y_\ell (\Omega_d) ) d\Omega_{d-2}^2 + L^2 d \Omega_d^2,
\end{align}
where $\omega$ is the quasinormal mode frequency. This fluctuation is subject to the linearized Einstein equations
\begin{equation}
\label{eq:linearized}
-\frac{1}{2} \left( \nabla_x^2 h_{\mu\nu} + \nabla_y^2 h_{\mu\nu} + \nabla_\mu\nabla_\nu \tensor{h}{^\rho_\rho} - \nabla^\rho\nabla_\mu h_{\rho\nu} - \nabla^\rho\nabla_\nu h_{\rho\mu} \right) + \frac{d-1}{L^2} h_{\mu\nu} = 0,
\end{equation}
where $\nabla_x^2$ and $\nabla_y^2$ are the d'Alembertians on AdS$_d$ and $S^d$ respectively and $\mu, \nu, \ldots$ index the $AdS_d$ coordinates. For the zero mode, $\omega=0$, we can drop $\chi_3$, and  it is straightforward to obtain the equations for the $\chi_i$'s. The gauge conditions fix $\chi_1$ and $\chi_4$, leaving a second order differential equation for $\chi_2$ coming from the $rr$ component of Einstein's equations
\begin{align}
\label{eq:chiw0}
-f \chi_2'' &+ \left[ \frac{ 2 d f^2 - 2 r^2 f'^2 + r f (-(d-2) f' + 2 r f'') }{r(rf'-2f)} \right] \chi_2'\\\nn
&+ \left[ \frac{2(d-1)r(rf'-2f)+ 2 L^2 d (ff' - rf'^2 + rff'')}{L^2 r(rf'-2f)}  \right]\chi_2 = -\frac{\ell(\ell+d-1)}{L^2} \chi_2.
\end{align}
There are higher order differential equations from the other components of Einstein's equations, but this equation is sufficient to satisfy all the others. We may call it the master equation \cite{Kodama:2003jz}. We can similarly proceed to derive the master equation for general $\omega$. We use the gauge conditions to fix $\chi_3$ and $\chi_4$, and then use the $rr$ Einstein equation to fix $\chi_1$. The rest of Einstein's equations can be manipulated to yield a single second-order differential equation for $\chi_2$, 
\begin{equation}
\label{eq:chi}
\chi_2'' + P \chi_2' + Q \chi_2 = 0,
\end{equation}
where $P$ and $Q$ are lengthy expressions in terms of $d, r,\omega^2, \ell, f$, and derivatives of $f$. We defer them to appendix \ref{app:chi2}. Since (\ref{eq:chiw0}) or more generally (\ref{eq:chi}) is the only equation to solve, we now drop the subscript on $\chi_2$.

In the analysis of quasinormal modes, one solves the $\chi$ equation (\ref{eq:chi}) with physical boundary conditions and obtains the dispersion relation $\omega = \omega(r_+, \ell)$. This relation contains information about how quickly 
a perturbation to the black hole (or equivalently the dual nonzero energy state of the field theory) dies off.   In particular, 
the zero mode, $\omega=0$, will give the threshold radius $r_*$ of instability. Since these quantities can be readily computed numerically, we shall obtain some insight into the quality of the large $d$ approximation for our setup. In addition, the analytic expression for $\omega$ will serve as a consistency check for the hydrodynamic equations that we derive later.

\subsection{Threshold radius and dispersion relation}
\begin{figure}
\begin{center}
            \includegraphics[width=4in]{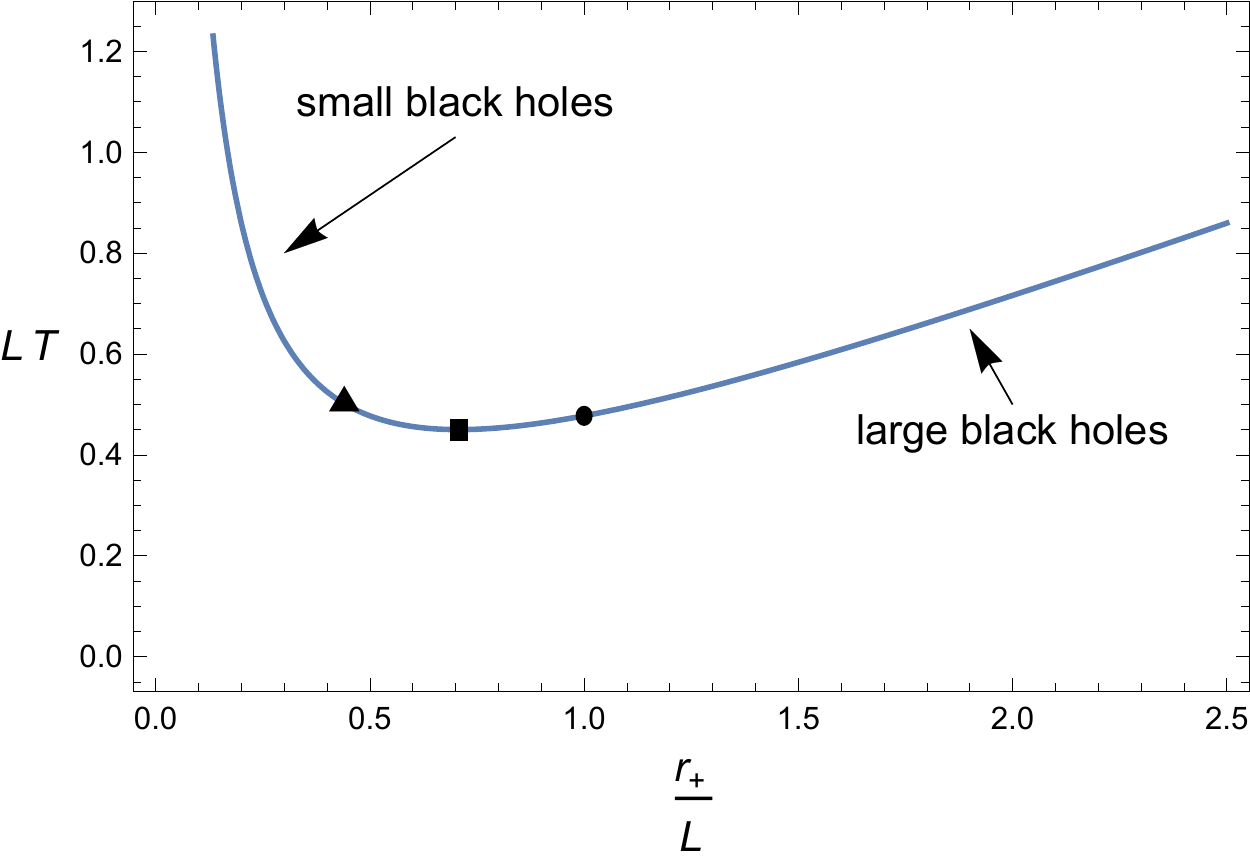}
\end{center}
 \caption{Black hole temperature as a function of radius for $d=5$.  
Approaching from the large black hole branch,  the circle indicates the Hawking-Page phase transition to thermal $AdS_5$ 
in the canonical ensemble, the square is the point at which the heat capacity becomes negative, and the triangle is the Gregory-Laflamme type instability \cite{Hubeny:2002xn}.  As $d$ gets larger, the points approach each other. }
\label{fig:largesmall}
\end{figure}

In an asymptotically $AdS_d$ spacetime with a $S^{d-2} \times {\mathbb R}$ conformal boundary (i.e.\ global coordinates), 
it is well-known that there are two branches of black hole solutions, small and large.  
Using the usual trick of analytic continuation into the Euclidean metric, the Hawking temperature is
\begin{equation}
T  = \frac{1}{4\pi} \left( \frac{(d-1)r_+}{L^2} + \frac{d-3}{r_+} \right).
\end{equation}
Fixing $T$, the resulting quadratic equation can be solved for two possible values of the horizon radius $r_+$ (see figure \ref{fig:largesmall}).
We pass three points of interest on this curve as we shrink the black hole radius down from infinity.   The first point we pass, at $(r_{\rm HP}, T_{\rm HP}) = (L, \frac{d-2}{ 2 \pi L})$, is the Hawking-Page phase transition.  This first order phase transition is to a thermal $AdS$ geometry 
\cite{Hawking:1982dh}, and is dual, through the AdS/CFT correspondence, 
to a confinement/deconfinement phase transition in the dual Yang-Mills theory \cite{Witten:1998zw}. 
The second point of interest,
\begin{equation}
r_{\rm HC}^2 = \frac{L^2 (d-3)}{d-1} \ , 
\end{equation}
 is where the small and large black hole branches meet and the heat capacity $C_v \sim dr_+/dT$ switches from a positive to a negative quantity.
In the limit $d \rightarrow \infty$, this critical radius is simply $r_{\rm HC} = L$, which is also where the Hawking-Page phase transition occurs. What we reconfirm in the analysis is that the dynamical, or Gregory-Laflamme like, instability at $r_+ = r_*$ sets in for slightly smaller black holes. In what follows, we choose $L=1$ which sets the distance scale.

To obtain the threshold radius for the dynamical instability, we need to examine the case $\omega=0$, for which the $\chi$ equation takes a considerably simpler form (\ref{eq:chiw0}). Due to the term $1/r^{d-3}$ in $f$, it is appealing to define $n \equiv d - 3$. 
The large $n$ limit needs to be taken carefully.  The gravitational effects of massive objects fall off very rapidly with distance at large $n$.  
To keep the black holes from completely decoupling, we follow \cite{Emparan:2013moa} and define a rescaled radial coordinate $R = (r/r_+)^n$.
The eigenvalues of the spherical harmonics diverge with $n$, and need to be rescaled.  Thus we define  $\hat{\ell} = \ell ( \ell + n + 2 ) / n$.

We impose the usual boundary conditions on $\chi$, i.e.\ $\chi$ is regular near the horizon $R\rightarrow1$, and decays sufficiently fast near the AdS boundary $R\rightarrow\infty$. Examining the equation (\ref{eq:chiw0}), there are two possible behaviors near the horizon, $\chi \sim 1$ and $\chi \sim 1/(R-1)$. A double power series in $(R-1)$ and then in $1/n$ gives for the first behavior
\begin{equation}
\chi \propto 1 + (R-1) \left[ -1 + \frac{1}{n} \frac{-6-4 r_+^2 + \hat{\ell} r_+^2}{2(1+r_+^2)} + \mathcal{O}(n^{-2})\right] + \mathcal{O} (R-1)^{-2}.
\end{equation}
At the zeroth order in $(R-1)$, it is simply constant with no higher order terms in $1/n$. This makes it easy to impose the horizon boundary condition, which is to eliminate all the terms of order $\sim \mathcal{O}{(R-1)^0}$ at all higher orders in $1/n$. 
There are also two modes near the AdS boundary, which scale as $\chi \sim 1$ and $\chi \sim 1/R$ at the leading order in $1/n$. The second mode can be expanded as
\begin{equation}
\chi \propto \frac{ R^{- \frac{4 + \hat\ell}{n} + \mathcal{O}(n^{-2})} }{R} \left(1 - \left(1 + \frac{\hat \ell}{2} + \mathcal{O}(n^{-1}) \right)R^{-2/n} + \mathcal{O}(R^{-4/n}) \right) 
\ .
\end{equation}
It is not entirely straightforward to match this expression to a series where we have first taken the large $n$ limit and then taken $R \to \infty$.  Further expanding each $R^{-2j/n}$ term above in a large $n$ limit will 
produce $\log(R)$ terms that need to be resummed.  
In what follows, we are able to apply boundary conditions simply by forbidding the $\chi \sim 1$ behavior and allowing for
$\log(R)^j / R$ terms, where $j$ is a small integer, in the large $R$ limit.

Given the boundary conditions, the most straightsforward way to solve the $\chi$ equation is to expand $\chi$ and $r_+$ as
\begin{equation}
\chi(R) = \sum_{j=0}^\infty n^{-j} \chi^{(j)}(R), \quad r_+(\hat \ell,n) = \sum_{j=0}^\infty n^{-j} r_+^{(j)}(\hat \ell).
\end{equation}
The differential equations at each order can be solved through DSolve in Mathematica \cite{Mathematica}. We can fix $r_+^{(j)}$ by imposing the boundary conditions. However, the equations get rapidly complicated, and it is difficult to go beyond first order for $r_+^{(j)}$. Instead, it was found useful to use the Green's method as laid out in \cite{Emparan:2013moa}, which allows us to move up one more order. At each order, we can write the $\chi$ equation in the self-adjoint form as
\begin{equation}
\frac{d}{dR}\left( R^2(R-1)^2 \frac{d\chi^{(j)}}{dR}\right) + 2R(R-1) \chi^{(j)} = S^{(j)},
\end{equation}
where
\begin{equation}
S^{(k)} = \sum^k_{j=1} -\mathcal{L}^{(j)}\chi^{(k-j)},
\end{equation}
and $\mathcal{L}^{(j)}$ is the differential operator in the master equation expanded in $1/n$ so that $\sum_j n^{-j} \mathcal{L}^{(j)}\chi = 0$. The solutions at each order are given by the Green's method as
\begin{eqnarray}
\chi^{(j)}(R) &=& A^{(j)} u(R) + B^{(j)} v(R) \\ \nn
&& + u(R) \int^\infty_R v(R')S^{(j)}(R')dR' + v(R) \int^R_1 u(R') S^{(j)}(R') dR',
\end{eqnarray}
where $u = \frac{1}{R}$ and $v = \frac{1}{1-R}$.
We then have the consistency condition on the sources $S^{(j)}$, which is derived from the Wronskian of $\chi^{(j)}$ \cite{Emparan:2013moa}
\begin{equation}
\lim_{R\rightarrow\infty} \frac1{R^2} \sum_j n^{-j} \int^RdR'u(R')S^{(j)}(R') = 0.
\label{eq:scond}
\end{equation}
This condition gives the same result as imposing the boundary condition near $R\rightarrow\infty$ on $\chi^{(j)}$, saving the effort to compute $\chi^{(j)}$ itself at the corresponding order. We were able to compute $S^{(j)}$ up to the fourth order, from which we find
\begin{equation}
r_*^2 = \frac{1}{1+\hat{\ell}} - \frac{1}{n} \frac{2 + \hat{\ell}}{(1+\hat{\ell})^2} + \frac{1}{n^2} \frac{4 + 5 \hat{\ell} + 2 \hat{\ell}^2}{(1+\hat{\ell})^3} + \mathcal{O}(n^{-3}).
\label{eq:thresh}
\end{equation}
This gives the threshold radius of the Gregory-Laflamme like instability to third order at large $n$. 
\newline
\newline
{\bf Dispersion relation for $\omega$ and $\ell$: \;} We can similarly proceed for the $\omega \neq 0$ case to obtain the dispersion relation, $\omega = \omega(\ell, r_+)$. The main difference is the appearance of waves near the horizon. Near the horizon at large $n$, the $\chi$ equation (\ref{eq:chi}) admits outgoing and infalling modes
\begin{equation}
\chi \propto (R-1)^{-1 + i\omega / 4\pi T}, \quad \chi \propto (R-1)^{-1 - i\omega / 4\pi T}.
\end{equation}
The infalling mode is conceived as physical, whose horizon expansion takes the following form
\begin{equation}
\chi \propto \frac{1}{R-1} \left[1 - \frac{1}{n} \frac{i r_+ \omega \log (R-1)}{1+r_+^2} + \mathcal{O}(n^{-2}) \right] + \left[ 
1 + 
\frac{i \hat{\ell} r_+}{\omega} + \mathcal{O}(n^{-1}) \right] + \mathcal{O} (R-1)^{-1}.
\end{equation}
As in the case of $\omega = 0$, at the AdS boundary, we only required that $\chi$ is suppressed by $1/R$.

Even with the Green's method, however, it is difficult to go beyond the second order for $S^{(j)}$, because the integral (\ref{eq:scond}) becomes hard to solve analytically. The leading order result is
\begin{equation}
\omega = i \left[ -(1 + \hat \ell)r_+ \pm \sqrt{ \hat \ell + ( 1 + \hat \ell )r_+^2 } \right].
\label{eq:disp}
\end{equation}
This gives the dispersion relation in the large $n$ limit, and yields the same leading order result for the critical radius (\ref{eq:thresh}).
At leading order in $n$, it makes no difference in the above expression if we write $\hat \ell$ or $\ell$.  In fact, the higher order corrections appear to be numerically smaller if we use $\ell$.  Furthermore, it is $\ell$ that naturally shows up in the nonlinear analysis we perform later.

\subsection{Comparison to numerical results}

For the purpose of numerical computation, it is useful to transform the original $\chi$ equation (\ref{eq:chi}) to simplify its asymptotic behavior. We introduce \cite{Buchel:2015gxa}
\begin{equation}
X(y) \equiv (1-y)^{1+i\omega / 4\pi T} y^{d+\ell+1} \chi(y), \quad y = \frac{r_+}{r}.
\end{equation}
In this variable and field, the quasinormal boundary conditions amount to
\begin{equation}
X( y \rightarrow 1 ) = 1, \quad X ( y \rightarrow 0 ) = \mbox{constant}.
\end{equation}

\begin{figure}
    \centering
    \begin{subfigure}[b]{0.48\textwidth}
        \includegraphics[width=\textwidth]{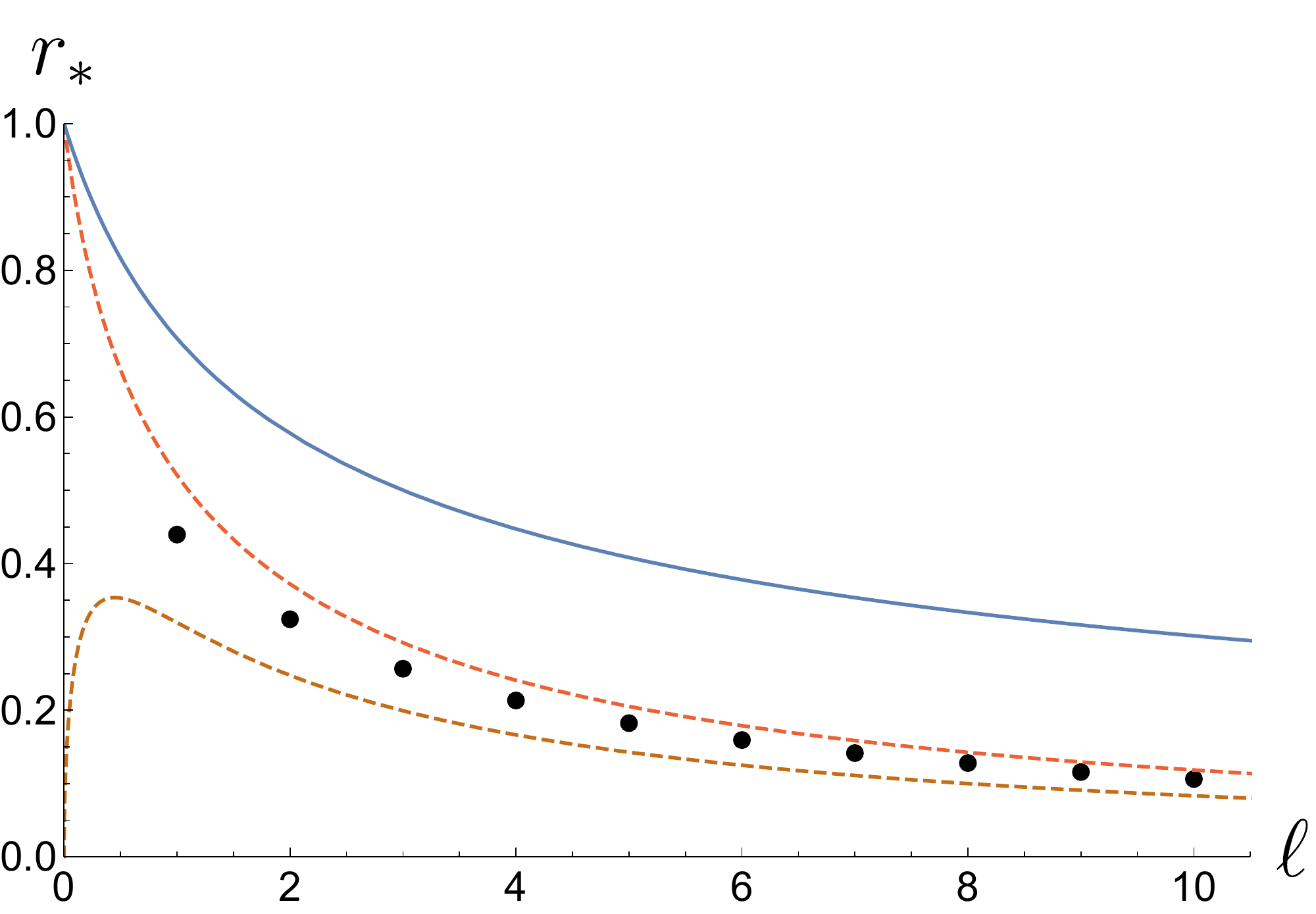}
        \caption{$d=5$}
        \label{fig:gull}
    \end{subfigure}
    ~ 
    \begin{subfigure}[b]{0.48\textwidth}
        \includegraphics[width=\textwidth]{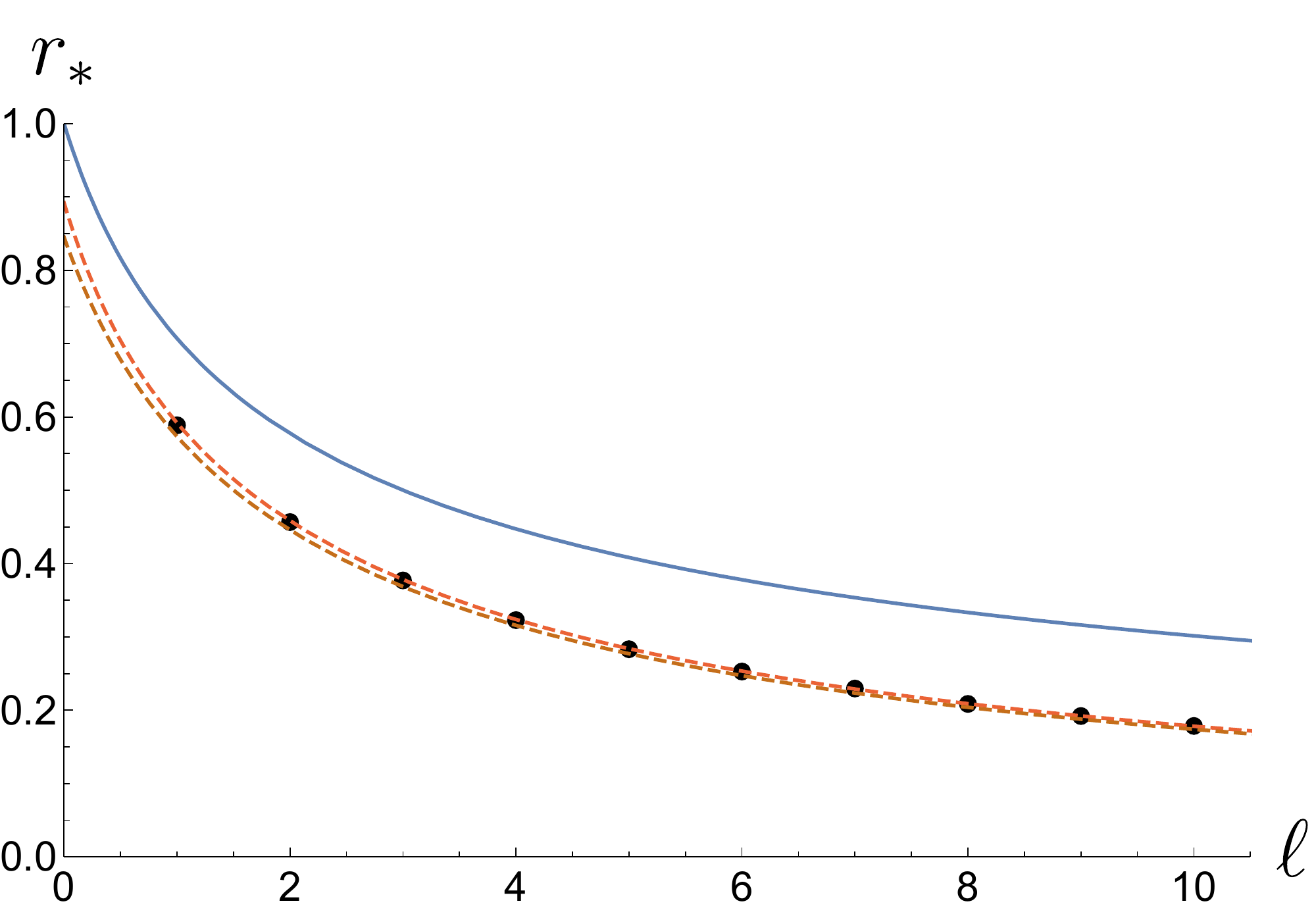}
        \caption{$d=10$}
        \label{fig:tiger}
    \end{subfigure}
\caption{Threshold radius of dynamical instability for $d=5, 10$. The dashed lines are the large $n=d-3$ approximation (\ref{eq:thresh}) including the terms up to $n^{-1}$ and $n^{-2}$ order, serving as the lower and upper bounds respectively. The dots are the numerical results; the solid line is the $n \rightarrow \infty$ limit.
}
\label{fig:thresh}
\end{figure}

The difference between the analytic formula for the threshold radius $r_*$ (\ref{eq:thresh}) and numerical calculation is near the order of expected error $\sim 1/n^3$, shown in fig.\ \ref{fig:thresh}.  Further, the two consecutive orders in the formula provide lower and upper bounds, at least for the range of $\ell$ we computed numerically. Both the analytic and numerical results indicate that the threshold of Gregory-Laflamme instability approaches asymptotically to $r_* \sim 1/\sqrt{1+\ell}$ for large $n$, reassuring us that the dynamical instability only develops below the thermodynamic threshold $r_* \sim 1$.

\begin{figure}
    \centering
    \begin{subfigure}[b]{0.48\textwidth}
        \includegraphics[width=\textwidth]{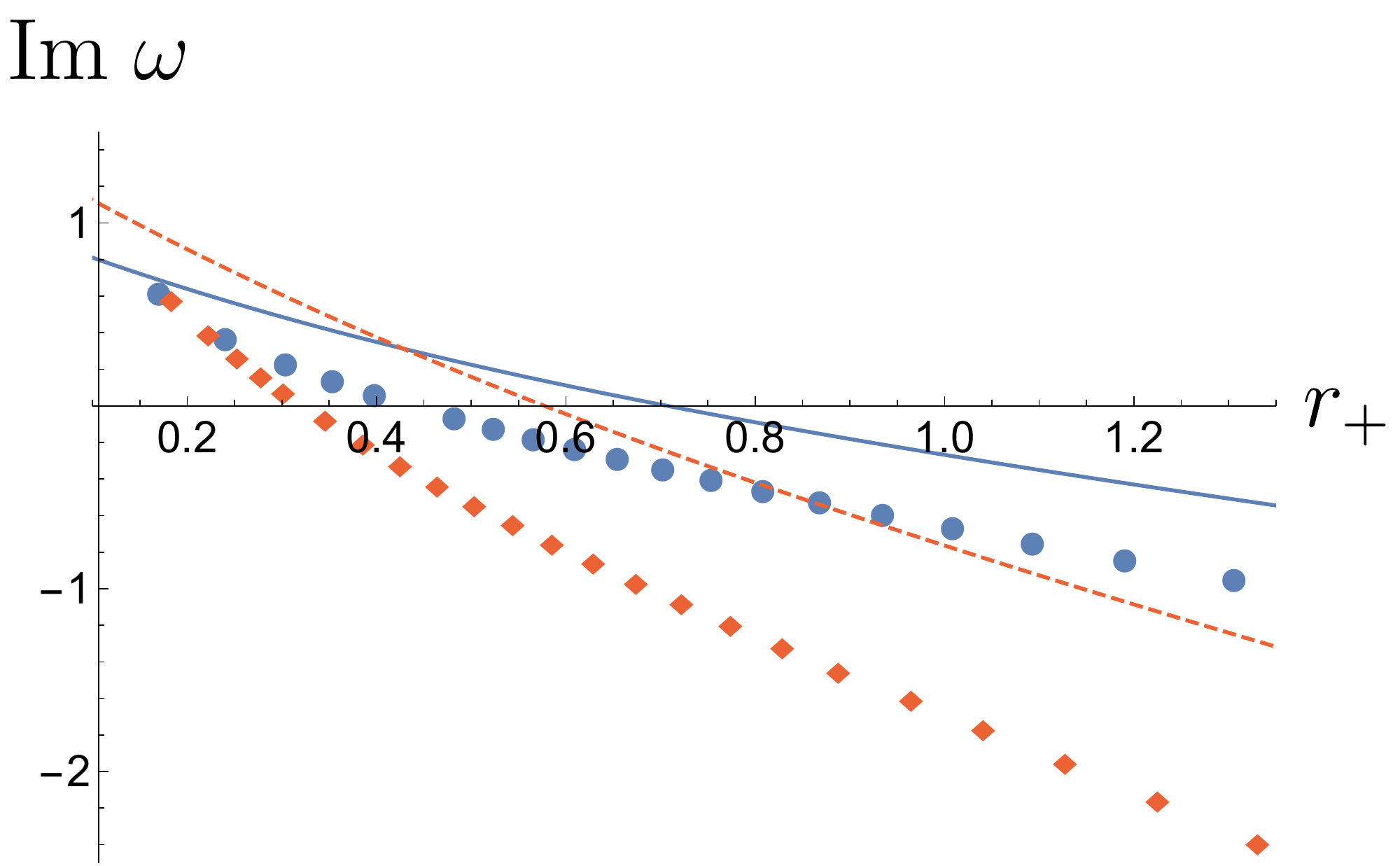}
        \caption{$d=5$}
        \label{fig:gull}
    \end{subfigure}
    \begin{subfigure}[b]{0.48\textwidth}
        \includegraphics[width=\textwidth]{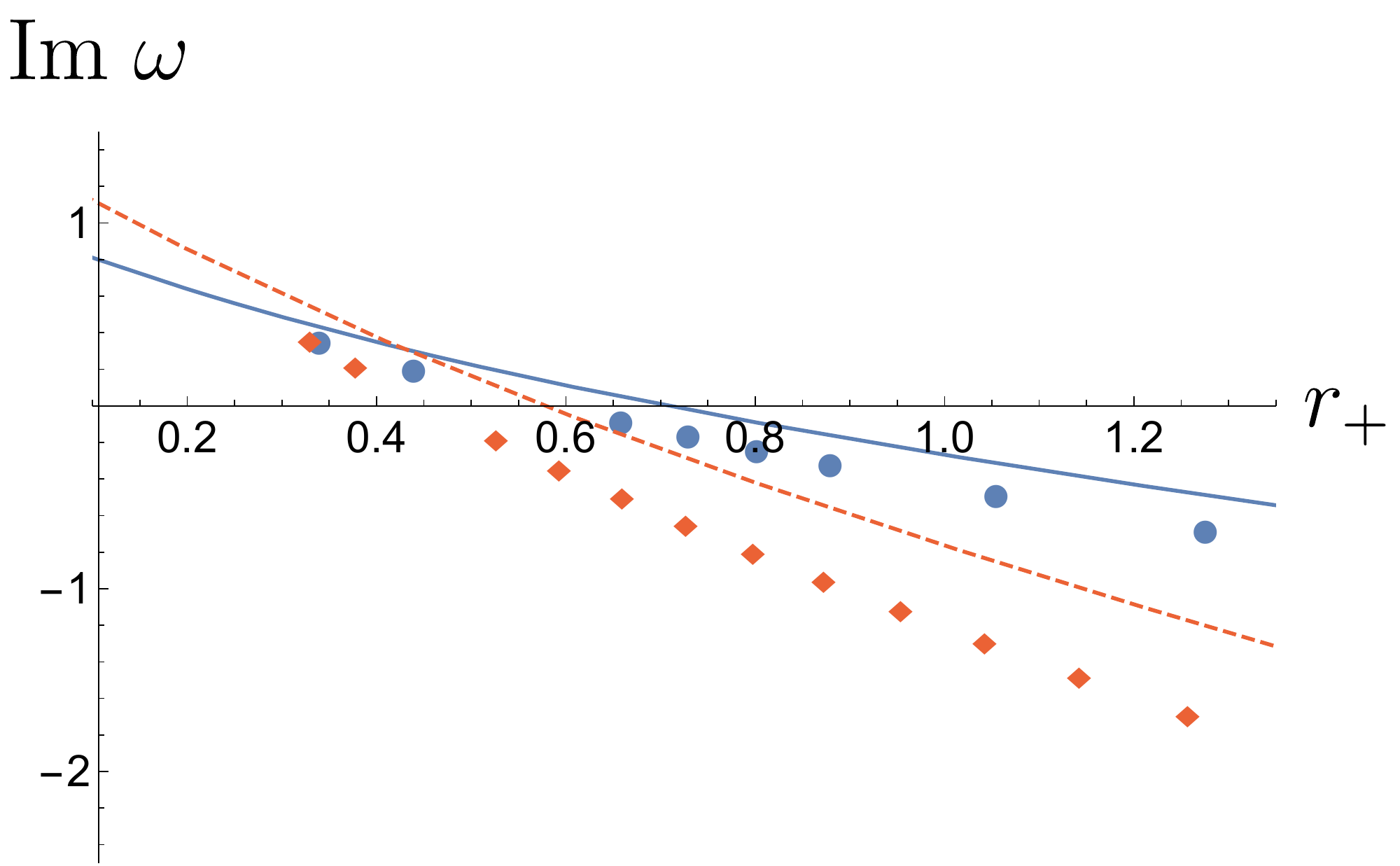}
        \caption{$d=10$}
        \label{fig:tiger}
    \end{subfigure}
\caption{Dispersion relation for $d=5, 10$. The solid and dashed lines are the large $n$ approximation (\ref{eq:disp}) for $\ell = 1,2$, respectively; the circle and diamond dots are the numerical results for $\ell = 1,2$, respectively.
To make these plots, we replaced $\hat \ell$ with $\ell$ in the relation (\ref{eq:disp}).
}
\label{fig:disp}
\end{figure}

The large $n$ approximation of the dispersion relation $\omega = \omega(r_+, \ell)$ (\ref{eq:disp}) also shows an expected deviation $\sim 1/n$ from the numerical calculation. In the numerical calculation, we assumed that $\omega$ is purely imaginary, which is manifest in the large $n$ limit (\ref{eq:disp}). As seen in fig.\ \ref{fig:disp}, although the deviation is not small for the dimension of interest, $d=n+3=5$, the approximation still captures the qualitative features.

\section{Evolution of black holes in the large $d$ limit}
\label{sec:3}

\subsection{Effective equations}

In the large $d$ limit, planar black holes are found to behave like a viscous fluid described by effective hydrodynamic-like equations \cite{Emparan:2016sjk}. A similar description is possible for our spherical black hole. To find such a description, 
we consider a metric and $d$-form ansatz of the following restricted form
\begin{equation}
\label{eq:metric}
ds^2 = g_{vv}dv^2 + 2dvdr + 2g_{v\theta}dvd\theta + g_{\theta\theta}d\theta^2 + g_A d\Omega_{d-2}^2 + g_B d\Omega_{d-1}^2,
\end{equation}
%
%
\begin{align*}
F_d = \;& \sqrt{g_A^{d-2}\det(S^{d-2})} \Big(f_{vr} dv \wedge dr \wedge d\Omega_{d-2} + f_{v\theta} dv \wedge d\theta \wedge d\Omega_{d-2} + f_{r\theta} dr \wedge d\theta \wedge d\Omega_{d-2} \Big)\\\nn
&+ \sqrt{g_B^{d-1}\det(S^{d-1})} \Big(f_v dv \wedge d\Omega_{d-1} + f_r dr \wedge d\Omega_{d-1} + f_\theta d\theta \wedge d\Omega_{d-1} \Big),
\end{align*}
where $g$'s and $f$'s are functions of $v$, $\theta$, and $r$ only.  
The coordinate $v$ is time-like, and $\theta$ is one of the coordinates on $S^d$.  
A key difference between this nonlinear and the previous linear analysis is that the $S^d$ here is allowed to fluctuate only with respect to the polar angle $\theta$.  
(Our metric is similar to the Kerr black hole, but we use the polar angle instead of the azimuthal angle.) 
We will restrict $F_d$ to be self-dual in accordance with supergravity.

Similar to the previous section, we use the scaled radial coordinate $R = (r/r_c)^n$, where $r_c$ is some reference scale and $n = d - 3$. We define the horizon radius $R_+$ to satisfy $g_{vv}(R_+) = 0$. In general, $R_+$ will depend on $v$ and $\theta$. However, by the relation $r_+ = r_c (R_+)^{1/n}$, the original horizon $r_+$ converges to $r_c$ in the large $n$ limit. Our large $n$ limit forces us to  probe the geometry close to the reference scale $r_c$, which is independent of $v$ and $\theta$.

In solving the Einstein equations (\ref{eq:eom}), it is convenient to exploit the fiber bundle structure of our metric ansatz. The Ricci tensor can be expressed in terms of the curvatures of the $S^{d-1}$ and $S^{d-2}$ plus additional contributions
(see the equations (\ref{eq:riccifull}) in the appendix \ref{app:ricci}). As to the boundary conditions, we require that the spacetime becomes $AdS_d \times S^d$ near the boundary $R \rightarrow \infty$ and regular near the horizon $R \rightarrow R_+$.\footnote{Note that we also expand the horizon radius as $R_+ = R_+^{(0)} + n^{-1} R_+^{(1)} + \dots$.} It is then tedious but straightforward to expand the $g$'s and $f$'s in $1/n$ and impose the boundary conditions. Some of the leading terms in the solution (see appendix \ref{app:metric}) are
\begin{align}
g_{vv} &= -\left(1 + r_c^2 R^{2/n} - \frac{e(v,\theta)}{R} \left( 1 + r_c^2 e(v,\theta)^{2/n} \right) \right) + \mathcal{O}(n^{-1}),\\\nonumber
g_{v\theta} &= \frac1n \frac{j(v,\theta)}{R} + \mathcal{O}(n^{-2}), \quad g_{\theta\theta} = 1 + \mathcal{O}(n^{-2}),\\\nonumber
g_A &= r_c^2 R^{2/n} + \mathcal{O}(n^{-3}), \quad g_B = \sin^2\theta  + \mathcal{O}(n^{-3}),\\\nonumber
f_v &= \mathcal{O}(n^{-2}), \quad f_r = \mathcal{O}(n^{-2}), \quad f_\theta = \sqrt{2(n+2)} + \mathcal{O}(n^{-3/2}).
\end{align}
We chose the field $e$ in such a way that $R_+ = e + \mathcal{O}(n^{-1})$. The fields $e$ and $j$ play the roles of fluid density and fluid velocity. They are subject to the constraints
\begin{align}
\label{eq:conserv1}
&\partial_v e - \cot \theta (r_c \partial_\theta e + j) = 0,\\
&\partial_v j - \cot \theta \left(r_c \partial_\theta j + \frac{j^2}{e} \right) + (1+r_c^2) \partial_\theta e + r_c (2+\csc^2\theta)j = 0.\nonumber
\end{align}
These hydro-like equations describe the evolution of the black hole in the large $d$ limit. They can be cast into a conservation law for a stress tensor on the hypersurface (with the normal $n = \partial_R$) near the AdS boundary,
\begin{equation}
\nabla_\mu T^{\mu\nu} = 0.
\end{equation}
The covariant derivative on the AdS boundary takes the form
\begin{align}
\nabla_\mu T^{\mu v} &= \partial_v T^{v v} + \partial_\theta T^{\theta v} + n T^{\theta v}   \cot\theta + \mathcal{O}(n^{-1}),\\\nonumber
\nabla_\mu T^{\mu \theta} &= \partial_v T^{v \theta} + \partial_\theta T^{\theta \theta} + n T^{\theta \theta} \cot\theta + \mathcal{O}(n^{-2}).
\end{align}
Then $e$ and $j$ constitute a stress tensor
\begin{align}
T^{\mu\nu} = 
\left(
 \begin{matrix}
  e + \mathcal{O}(n^{-1}) & n^{-1}(-j - r_c \partial_\theta e) + \mathcal{O}(n^{-2})\\
  n^{-1}(-j - r_c \partial_\theta e) + \mathcal{O}(n^{-2}) & T^{\theta\theta}
  \end{matrix}\right)
\label{eq:stress}
\end{align}
where
\begin{align}
T^{\theta\theta} &= \frac{1}{n^2} \left[ -2 r_c(\csc\theta \sec\theta + \tan\theta) j + \frac{j^2}{e} \right. \\\nonumber
&\quad \left. - \left(r_c^2 \csc\theta \sec\theta +(1+r_c^2) \tan\theta \right) \partial_\theta e + 2 r_c \partial_\theta j + r_c^2 \partial_\theta^2 e \vphantom{\frac{1}{n^2}} \right] + \mathcal{O}(n^{-3}).
\end{align}
This form is very similar to that in \cite{Herzog:2016hob}. 
We are not claiming that this stress tensor is the stress tensor of a dual field theory through the usual AdS/CFT dictionary.  In fact, given the internal $\theta$ index, such an interpretation is obscure at best.  
\newline
\newline
{\bf Zooming in on the equator:} One troubling aspect of the conservation equations (\ref{eq:conserv1}) is that $\partial_\theta T^{v\theta}$, or equivalently $\partial_\theta j$, 
does not appear at leading order in the $1/n$ expansion. 
We can restore such a $\theta$ derivative by rescaling $\theta$ and $j$:
\begin{equation}
j \rightarrow j' = \frac{j}{\sqrt{n} }, \quad \theta \rightarrow u = \sqrt{n} \left( \theta - \frac{\pi}{2} \right).
\end{equation}
In the large $n$ limit, this rescaling zooms in on the equator of the $S^d$, where $\theta \sim \pi/2$. 
In refs.\ \cite{Emparan:2015gva, Emparan:2016sjk,  Herzog:2016hob}, a similar rescaling of the distance ruler was motivated in part by a desire to keep the sound modes in the fluctuation spectrum.  For a conformal fluid, the speed of sound scales as $1 / \sqrt{n}$.  By changing the distance ruler, the speed of sound is kept finite rather than infinitesimal, and a spatial gradient of $j$ reappears in the energy  conservation equation. 
In our case, the sound modes are gapped by the curvature of the sphere, but a spatial gradient of the current is restored in a similar way.

Solving the equations of motion order by order with this new scaling, we find
%
\begin{align}
\label{eq:conserv2}
&\partial_v e - \partial_u (j + r_c \partial_u e) + u (j + r_c \partial_u e) = 0,\\
&\partial_v j - \partial_u \left( r_c \partial_u j + \frac{j^2}{e} \right) + u \left( r_c \partial_u j + \frac{j^2}{e} \right) + r_c (j + r_c \partial_u e) + \partial_u e + 2 r_c j = 0.\nn
\end{align}
Unlike the previous effective equations (\ref{eq:conserv1}), we were not able to recast these equations in the form of a conservation law for a stress tensor. 
Nonetheless, the rescaled equations have the virtue of taking a derivative of the momentum-like quantity $j + r_c \partial_u e$. They resemble the equations for the black strings \cite{Emparan:2015gva}, which is perhaps not surprising as stretching the polar angle $\theta$ makes the sphere look like a string.

\subsection{Solving the Unscaled System}
\label{sec:solution}

\begin{figure}
    \centering
    \begin{subfigure}[b]{.55\textwidth}
        \includegraphics[width=\textwidth]{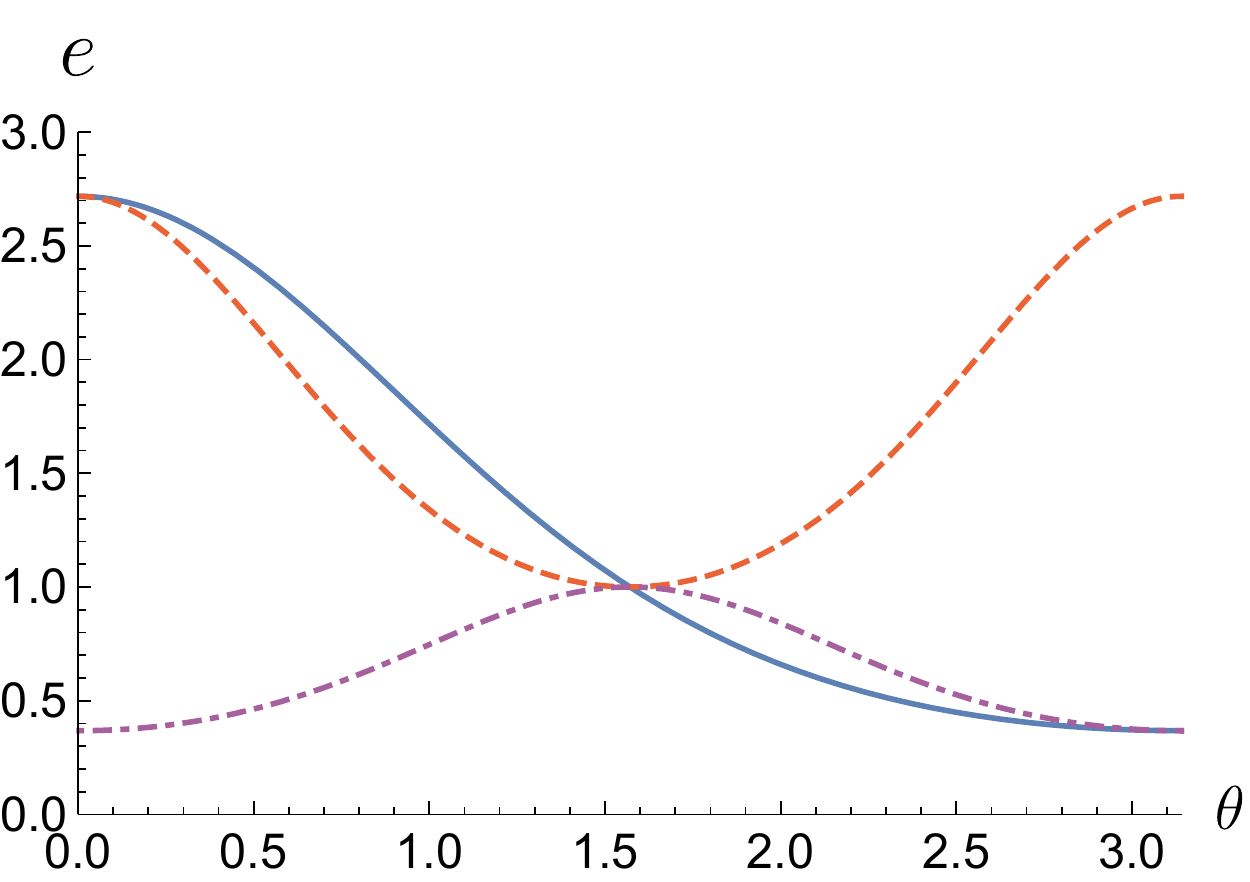}
    \end{subfigure}
\caption{Snap shots of solutions (\ref{egeneral}) that lump near poles and form a belt around the equator. The single-spot solution (solid) corresponds to $\ell = 1$
and $a_{1+} > 0$ (with all other $a_{\ell \pm} = 0$). The double-spot solution and belt solution (dashed and dot-dashed) correspond to $\ell = 2$ and $a_{2+} = 1$ or $-1$ (with all other $a_{\ell \pm} = 0$). 
}
\label{fig:steady}
\end{figure}

\begin{figure}
    \centering
    \begin{subfigure}[b]{\mywidthb\textwidth}
        \includegraphics[width=\textwidth]{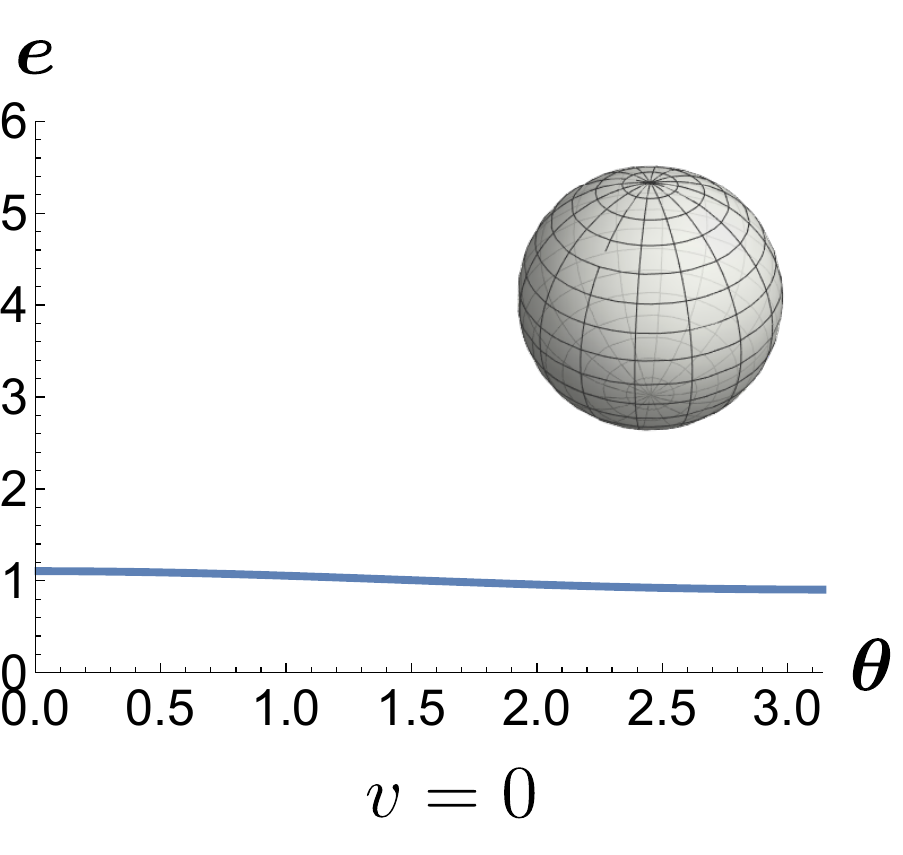}
    \end{subfigure}
    \begin{subfigure}[b]{\mywidthb\textwidth}
        \includegraphics[width=\textwidth]{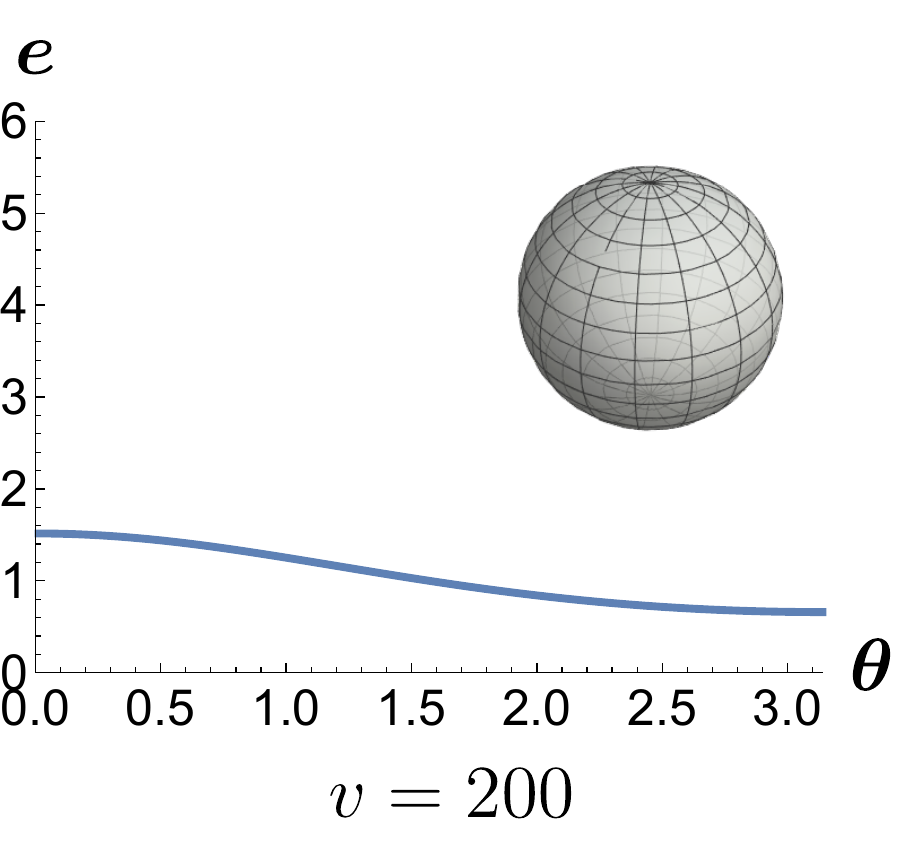}
    \end{subfigure}
    \begin{subfigure}[b]{\mywidthb\textwidth}
        \includegraphics[width=\textwidth]{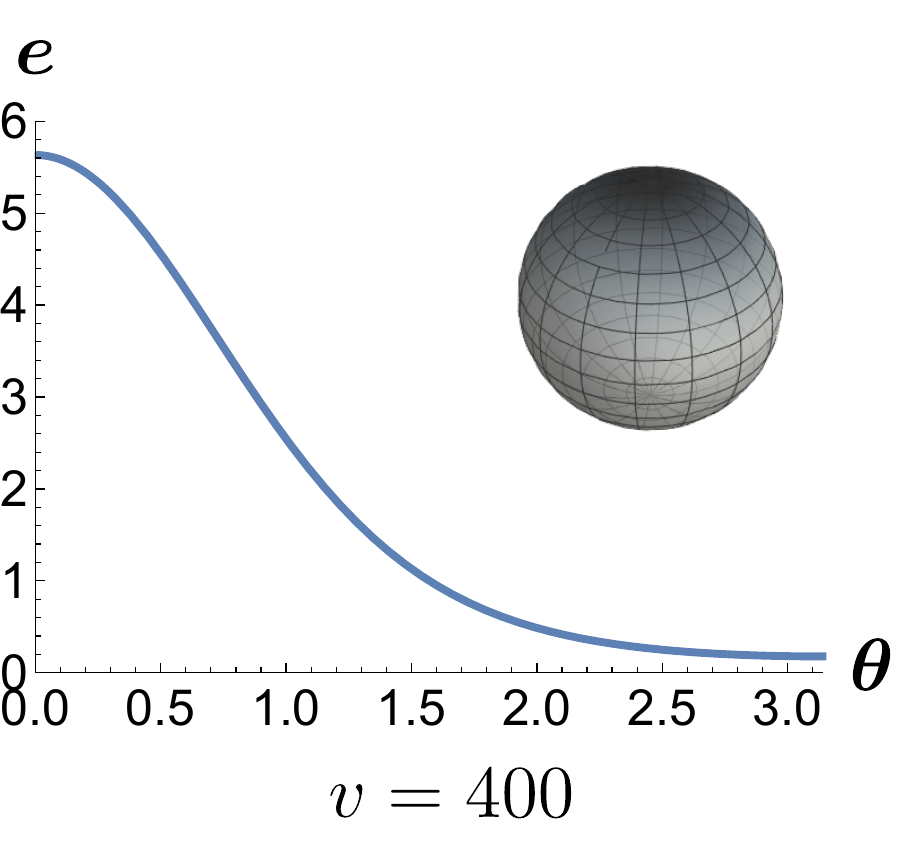}
    \end{subfigure}
    \begin{subfigure}[b]{\mywidthb\textwidth}
        \includegraphics[width=\textwidth]{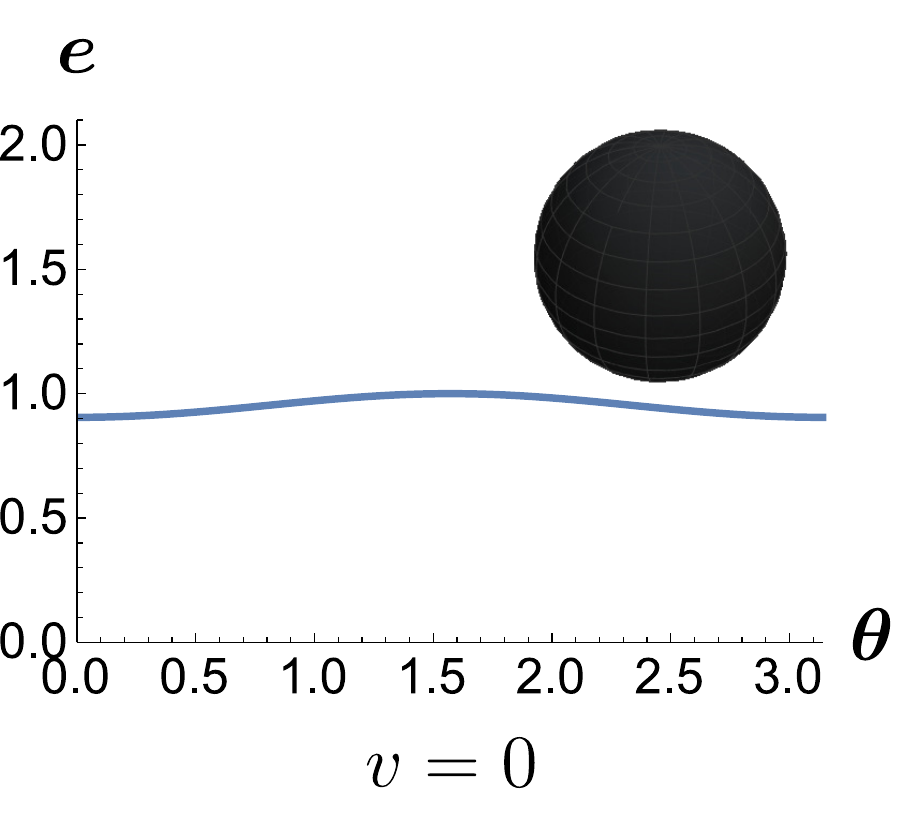}
    \end{subfigure}
    \begin{subfigure}[b]{\mywidthb\textwidth}
        \includegraphics[width=\textwidth]{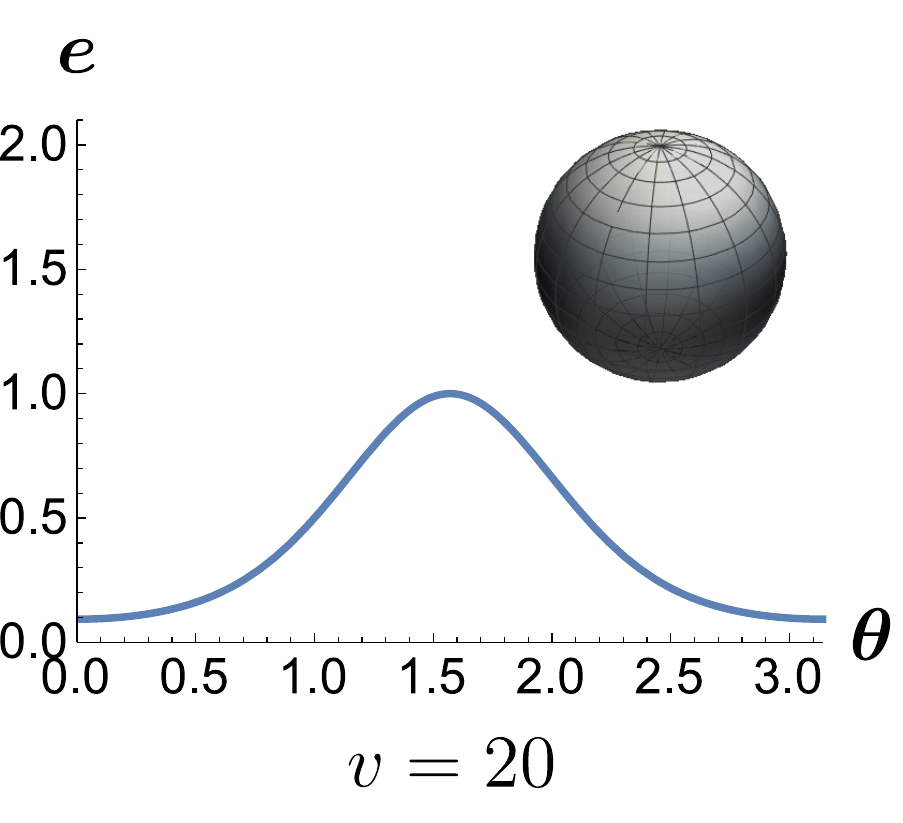}
    \end{subfigure}
    \begin{subfigure}[b]{\mywidthb\textwidth}
        \includegraphics[width=\textwidth]{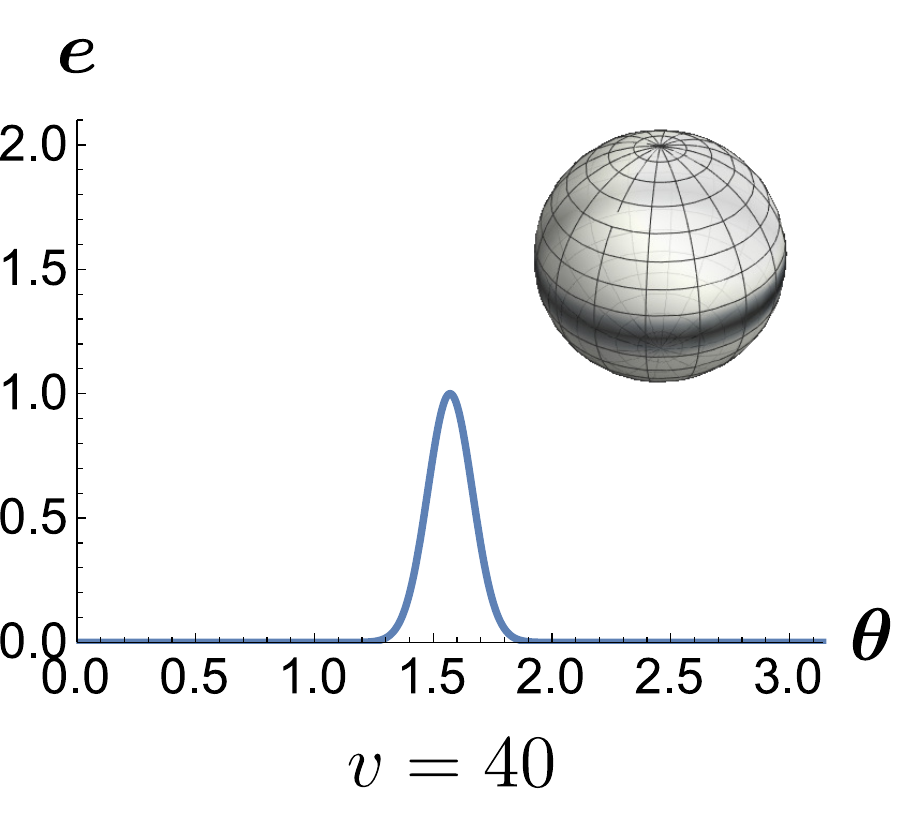}
    \end{subfigure}
    \caption{Snap shots of solutions (\ref{egeneral}) at different times. The almost uniform black holes evolve into the non-uniform black holes, but there is no end-point. We display the density plot on the sphere in insets. {\bf First row:} We used the $\ell=1$ term with  $a_{1+} > 0$ (all other $a_{\ell\pm}=0$), and set the radius $r_c=0.7$. {\bf Second row:} We used the $\ell=2$ term with $a_{2+} < 0$ (all other $a_{\ell\pm}=0$), and set the radius $r_c=0.5$.
}
\label{fig:evol1}
\end{figure}

We were able to find a large class of solutions to the ostensibly nonlinear equations (\ref{eq:conserv1}).  
In fact, taking a logarithm of $e(v, \theta)$ leads to a linearization and an ability to superpose different solutions.
The first equation can be solved to give an expression for $j(v, \theta)$ in terms of $e(v,\theta)$:
\begin{eqnarray}
j = -r_c \partial_\theta e + \tan \theta \, \partial_v e \ .
\end{eqnarray}
The following expression for $e(v, \theta)$ then gives a solution of the coupled system:
\begin{eqnarray}
\label{egeneral}
e = c \exp \left( \sum_{\ell = 1}^\infty \left( a_{\ell +} e^{-i \omega_{\ell +} v} +
a_{\ell -} e^{-i \omega_{\ell -} v} \right) \cos^\ell \theta \right) \ ,
\end{eqnarray}
where $a_{\ell \pm}$ and $c$ are integration constants and the frequencies obey the dispersion relations
\begin{equation}
\omega_{\ell \pm} = i \left[ - r_c (\ell+1) \pm \sqrt{\ell + r_c^2 (\ell+1)} \right] \ .
\end{equation}
This dispersion relation is precisely what we found earlier (\ref{eq:disp}) for the quasinormal modes driving the instability.
One choice of sign $\omega_{\ell-}$ always leads to a damped response.  However, $\omega_{\ell +}$ can lead to an exponentially 
growing response when the radius is below the threshold for the corresponding instability, $r_c^2 \leq \frac{1}{\ell + 1}$.  
In figures \ref{fig:steady} and \ref{fig:evol1}, we plot snap-shots of these time evolving solutions for $\ell=1$ and 2.

For small $a_{\ell \pm}$, we can expand out the exponential and consider the time dependence perturbatively.  
The quasinormal modes we considered before were constructed from spherical harmonics $Y_{\ell} (\Omega_d)$.  The $\cos^\ell \theta$ dependence we find here is in fact a large $d$ limit of one of these spherical harmonics.  
As our metric ansatz depends only on the polar angle $\theta$ of the $S^d$, we should be careful to select those spherical harmonics which have only such a $\theta$ dependence.  
Such spherical harmonics exist, and in the large $d$ limit, they simplify to $\cos^\ell \theta$ (see appendix \ref{app:spher}).

Detailed numerical calculations in the $d=5$ case \cite{Dias:2015pda, Dias:2016eto} indicate the existence of static black hole solutions that break the $SO(6)$ symmetry of the $S^5$ down to $SO(5)$.  The black hole horizon radius becomes a function of the polar angle $\theta$ on the $S^5$.  In limiting cases, 
the black hole radius can even shrink to zero for a range of the $\theta$-parameter, yielding black belt and spot solutions where the topology of the black hole changes to $S^4 \times S^4$ and $S^8$ respectively.
While our solutions (\ref{egeneral}) are in general strongly time dependent, they share some qualitative features with these spot and belt solutions.  The odd $\ell$ solutions lead to single spots at the north or south pole, depending on the sign of $a_{\ell \pm}$.  The even $\ell$ solutions in contrast lead to either belt solutions or solutions with two spots -- one at the north and one at the south pole -- depending again on the sign of $a_{\ell \pm}$.  

We cannot make $e(v, \theta)$ arbitrarily small (or large) and stay within the regime of validity of our large $d$ approximation.  We need to keep 
$|\log e(v,\theta)|  \ll n$.  Thus, we cannot strictly speaking witness a topology change in the shape of the black hole horizon.  Nevertheless, the exponential growth and/or suppression in the solution (\ref{egeneral}) is very suggestive that such endpoints may exist, that there exist black spot and belt solutions with topology $S^{2d-2}$ and $S^{d-1} \times S^{d-1}$ respectively.  One might conjecture that subleading terms in the $1/d$ expansion could halt the exponential growth of the solutions at late times, and also possibly lead to topology changes in the horizon.    

At precisely the instability threshold $\omega_{\ell+}=0$ for integer $\ell$, 
the corresponding $\cos^\ell \theta$ mode can lead to a inhomogeneous static solution. 
Naively, for a general $r_c$ there exists a non-integer $\ell$ for which $\cos^\ell \theta$ also leads to a inhomogeneous static solution.  The issue with such solutions is that one needs an absolute value sign, $|{\cos \theta} |^\ell$, to keep the solution real on the whole sphere, and then there are corresponding non-analyticies in the solution at the equator and distributional terms which do not cancel out of the equations of motion.

\subsection{Solving the Rescaled System}
We have not been able to find general time dependent solutions of the rescaled hydrodynamic like equation (\ref{eq:conserv2}).
We first investigate steady state solutions of the rescaled system.
If we assume no time independence and make the ansatz
\begin{equation}
j_s(u) = - r_c \partial_u e_s(u) \ ,
\end{equation}
then static configurations for $e(u)$ must satisfy the following nonlinear equation:
\begin{equation}
f''(u) + f(u) \left( -2 + \frac{1}{r_c^2} + f'(u) \right) - u f'(u) = 0 \ ,
\end{equation}
where $f(u) \equiv e'(u) / e(u)$ and $'$ denotes $\partial_u$.  One very simple solution is a Gaussian configuration where $f(u) = -( r_c^{-2} -3)  u$ and
\begin{equation}
\label{rescaledgaussian}
e_s(u) = e_0 \exp\left( - \left( \frac{1}{r_c^2} - 3 \right) \frac{u^2}{2} \right)   \ .
\end{equation}
In order for this solution to damp out as $|u|$ gets large, we require $r_c^2 < \frac{1}{3}$, which is the stability threshold for the $\ell = 2$ mode.
Because this static solution is symmetric under $u \to - u$, it also exists on a hemisphere where we quotient out by this reflection symmetry about the equator, eliminating all of the odd parity modes, including the most unstable $\ell = 1$ mode.  It is therefore 
sensible that it is the $\ell =2$ mode which determines whether this static solution can exist or not.\footnote{%
We have found a couple of additional static solutions at discrete values of the horizon radius $r_c$.  At $r_c^2 = 1/2$, corresponding to the onset of the $\ell=1$ instability, an exponential profile $e(u) \sim \exp(c u)$ is a solution.  At $r_c^2 =1$, there is a quadratic solution $e(u) \sim u^2$.  
}

We can then further look at fluctuations about the Gaussian solution.  We look for fluctuations of the form
\begin{align}
 e(v, u) &= e_s(u) + e^{-i v \omega} \delta e(u)  \ , 
\\
j(v, u) &=  j_s(u) + e^{-i v \omega} \delta j(u) \ .
\end{align}
We find solutions
\begin{align}
\delta e(u) &=  e^{ u^2/2}   \partial_u f(u) \ ,
\\
\delta j(u) &= - r_c \partial_u \left( e^{u^2/2} \partial_u f(u) \right) - i \omega e^{u^2/2} f(u) \ ,
\end{align}
where
\begin{align}
f(u) &= c \, H_q \left( \left( \frac{1}{2 r_c^2} - 1\right)^{1/2}  u \right) \exp \left( -\left(\frac{1}{r_c^2} - 2 \right) \frac{u^2}{2} \right) \ ,
\\
0 &= \omega^2 - i \left( 4(q-1) r_c - \frac{2q}{r_c} \right)\omega  - \frac{(q+1)(2r_c^2-1)(r_c^2 + q(2r_c^2-1))}{r_c^2} \ ,
\end{align}
and where $H_q(x)$ is a Hermite polynomial and $c$ an arbitrary constant.  There are zero modes when
\begin{equation}
r_c = \frac{1}{\sqrt{2}} \ , \sqrt {\frac{q}{1+2q}} \ .
\end{equation}
In between these points, the frequency goes into the upper half plane, and the mode exponentially grows in time.  The upper threshold for stability is  the same as that for the $\ell=1$ mode, $r_c^2 = 1/2$.  Interestingly, the lower threshold $\sqrt{\frac{q}{1+2q}}$ is always greater than or equal to $1/\sqrt{3}$.  We need $r_c^2 < 1/3$ to have the Gaussian background solution, and so these fluctuations appear to be stable with respect to the Gaussian background.

We now investigate the full, nonlinear time evolution of these black hole solutions in the large $d$ limit and in the rescaled variables.
The global effective equations 
near the equator (\ref{eq:conserv2}) are simple enough that 
they can be
solved numerically with prepackaged differential equation solvers, such as the NDSolve function in Mathematica \cite{Mathematica}.

\begin{figure}
    \centering
    \begin{subfigure}[b]{\mywidthc\textwidth}
        \includegraphics[width=\textwidth]{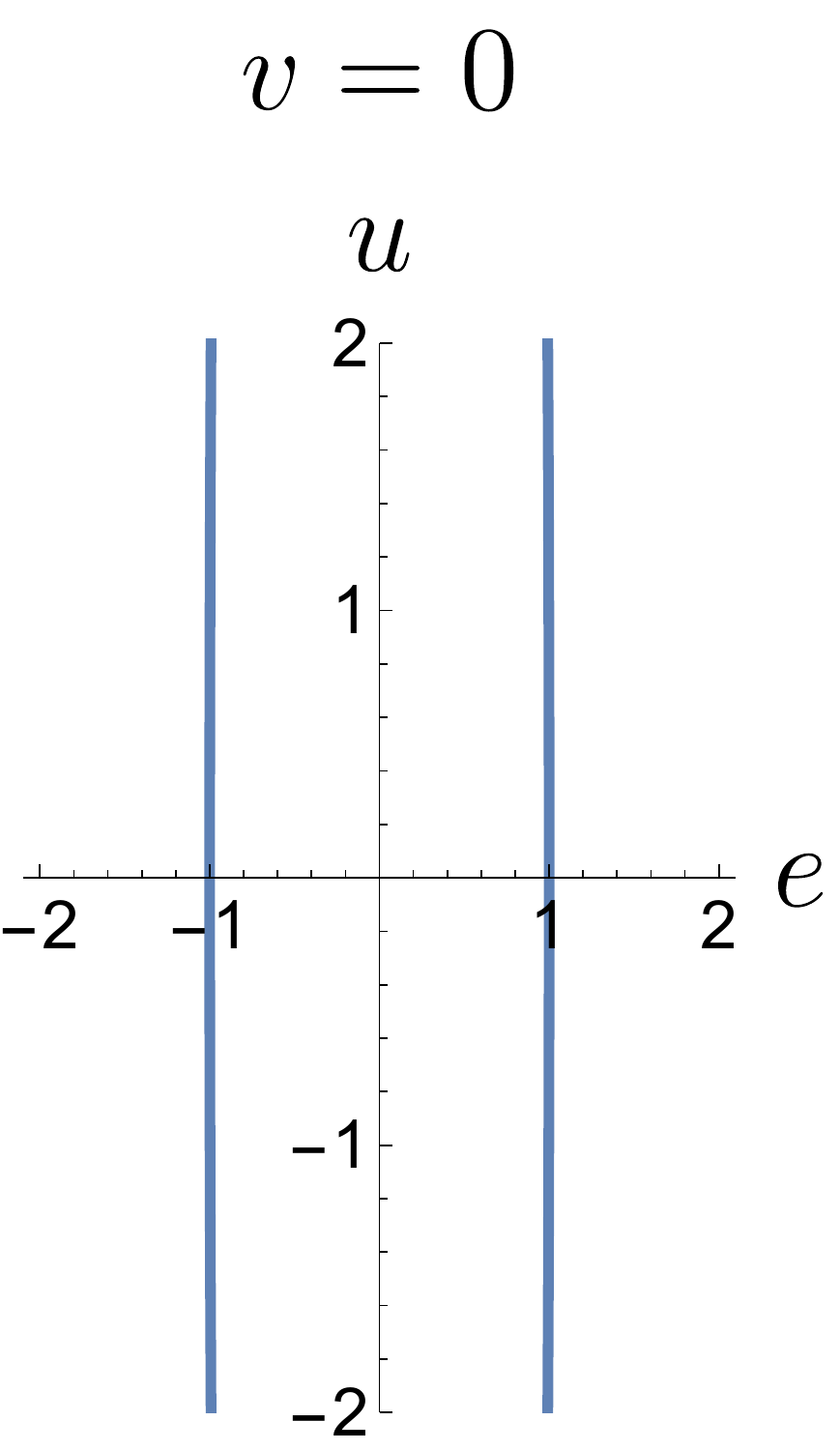}
    \end{subfigure}
    \begin{subfigure}[b]{\mywidthc\textwidth}
        \includegraphics[width=\textwidth]{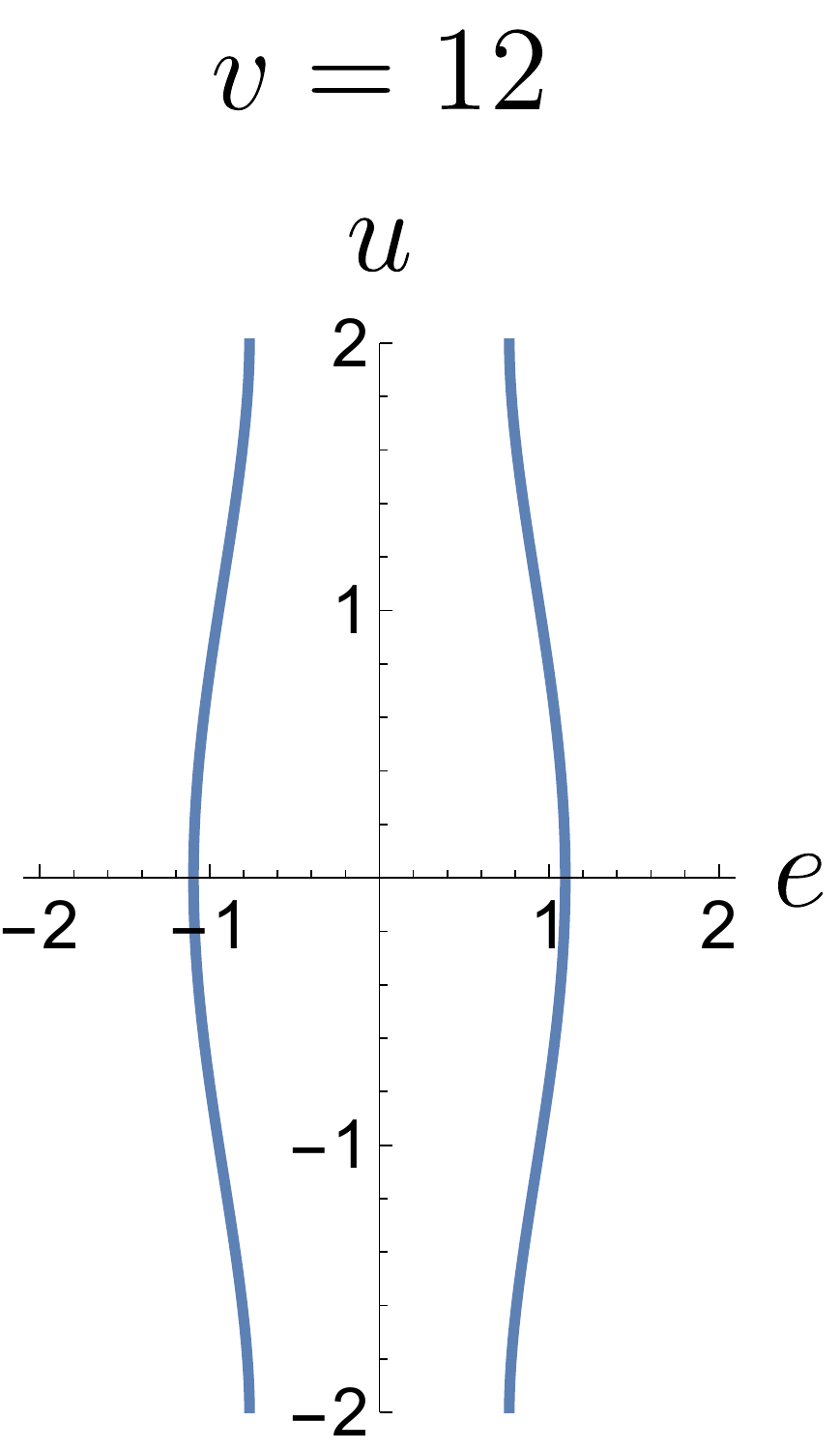}
    \end{subfigure}
    \begin{subfigure}[b]{\mywidthc\textwidth}
        \includegraphics[width=\textwidth]{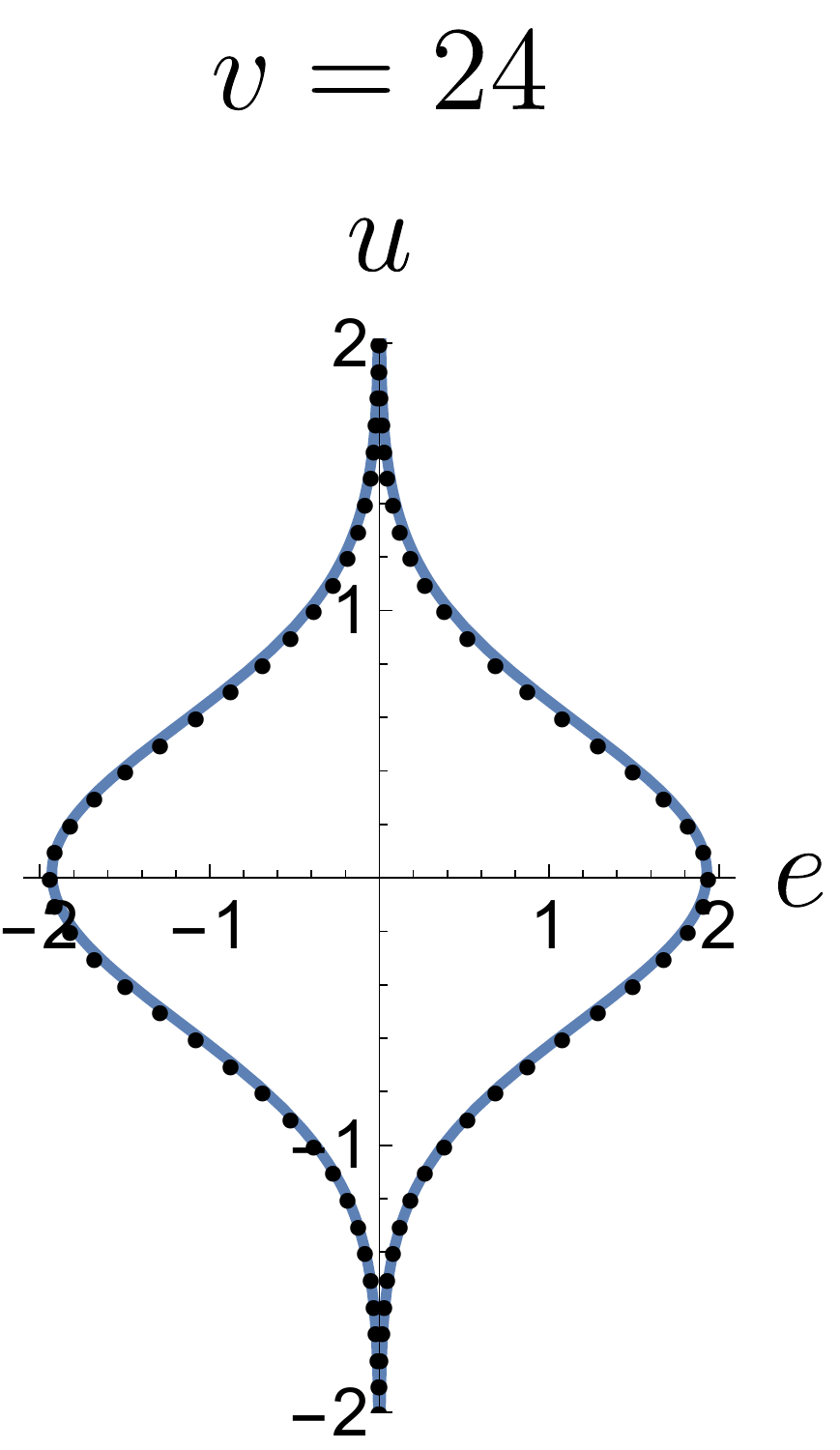}
    \end{subfigure}
    \begin{subfigure}[b]{\mywidthc\textwidth}
        \includegraphics[width=\textwidth]{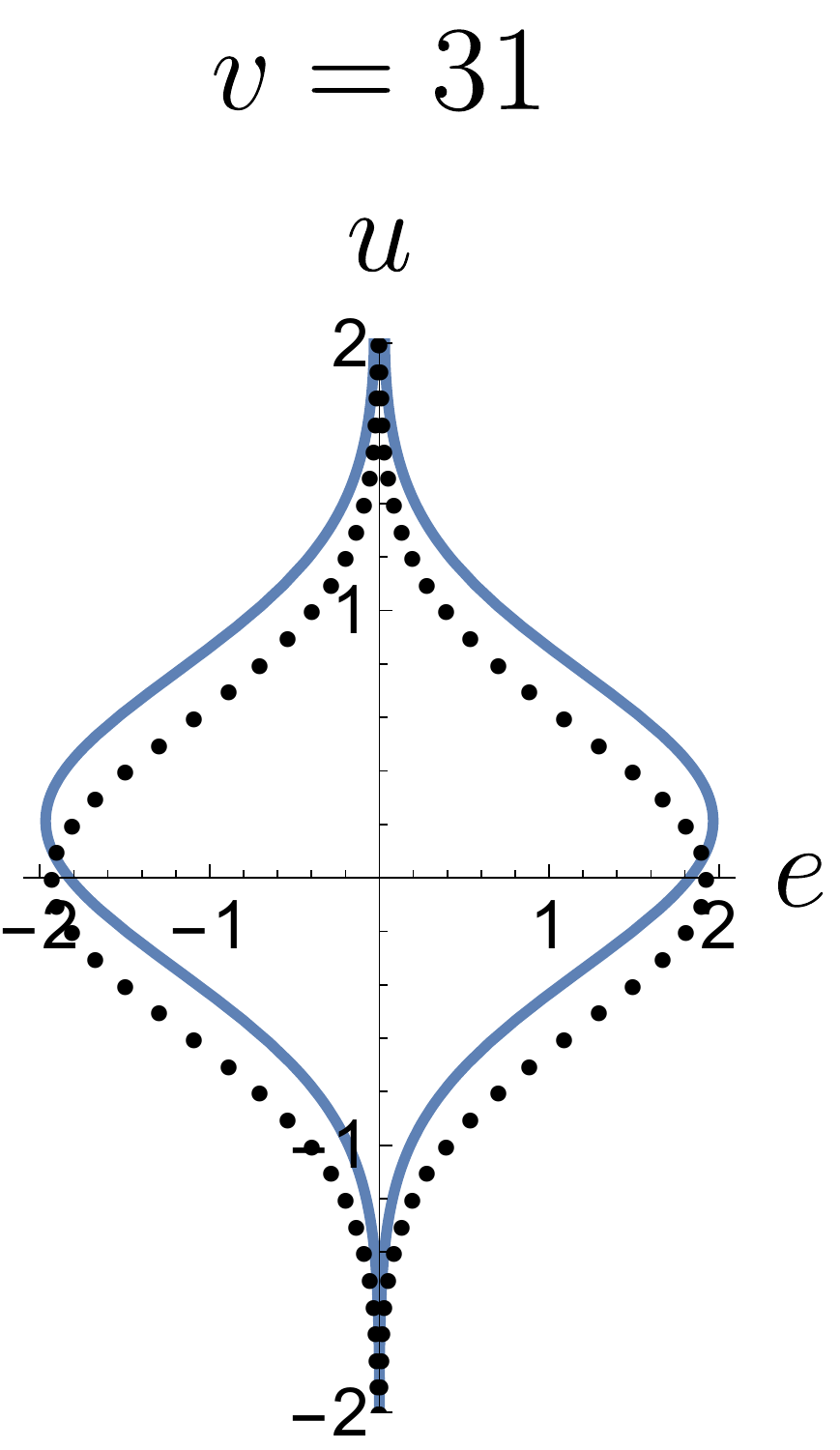}
    \end{subfigure}
    \caption{Response of the uniform black hole to the perturbation $\delta e = \epsilon u^2$ with $\epsilon < 0$ for the rescaled equations (\ref{eq:conserv2}). The reference radius of the black hole is $r_c=0.4$. For the last two plots, we overlap the steady gaussian solution (\ref{rescaledgaussian}) as dotted lines. The amplitude $e_0 \sim 1.94$ was fitted to the solution at $v=24$. This gaussian state moves until NDSolve crashes shortly after  $v = 31$.}\label{fig:evol3}
\end{figure}

For these equations, provided we begin with a black hole radius $r_c^2 < 1/3$ with a perturbation that preserves the $u \to -u$ symmetry, 
we can evolve to an approximate steady-state that closely matches the Gaussian profile 
(\ref{rescaledgaussian}).  See fig.\ \ref{fig:evol3} for one such numerical evolution.  
We say approximate because we have observed a different type of instability that kicks in at long times, where the Gaussian profile starts to move toward one pole or the other, spontaneously breaking the parity symmetry.  We suspect this instability is a remnant of the $\ell=1$ instability we saw in the original $\theta$ variable, but we have not been able to find an analytic form for it in the rescaled case.
In the regime $1/3 < r_c^2 < 1/2$, we do not find any approximate steady-state solutions that resemble Gaussians.  Indeed, our earlier analysis indicated the Gaussian solution only exists for $r_c^2 < 1/3$.  Instead, the numerics suggests evolution toward a spot-like solution at the north or south pole.  However, the time evolution crashes before a stationary solution is reached, at least in part because of the exponentially growing nature of the energy density at the poles.
For $r_c^2 > 1/2$, perturbations to the black hole solution die out, and the $SO(d+1)$ symmetric solution appears to be stable.  

The rescaled equations (\ref{eq:conserv2}) are second order rather than first order in the spatial derivative, necessitating the introduction of a second pair of boundary conditions.  In addition to the Dirichlet condition on the current $j=0$ at the poles, we introduce a Neumann condition on the energy density, that $\partial_\theta e = 0$.  
These boundary conditions imply that the integral of $e$ over the $\theta$ direction is a conserved quantity.

There are some technical issues with these rescaled coordinates.
The points $u \to \pm \infty$ are singular and we can only solve the differential equations in some range $-L < u < L$.   In practice, the boundary conditions are not applied at the poles but at $u = \pm L$.  
 A further difficulty is that with the Gaussian type solutions we find for $e$, the ratio $j^2 / e$ in the differential equations can only be evaluated accurately provided $L$ is not too big.  
 A final issue here is that at any fixed and finite $n$, once $u \sim \sqrt{n}$, we need to include subleading terms in our large $n$ expansion.

\subsection{Entropy and Conservation of Energy}

We examine energy and entropy constraints.
From the first of the evolution equations (\ref{eq:conserv1}), it follows that the following integral over the energy density is time independent, to leading order in $n$:
\begin{equation}
E_{\rm tot} \equiv \int_0^\pi e(v, \theta)  \sin^n \theta \, d \theta \ .
\end{equation}
The $\sin^n \theta$ factor is a bit like a Dirac delta function, peaked at $\theta = \pi/2$.   The behavior of $e(v, \theta)$ at the poles is thus only very weakly constrained by energy conservation in this large $n$ limit.  In this light, it is perhaps less surprising that the time evolution discussed above sometimes leads to diverging values of $e(v,\theta)$ near the poles.  On the other hand, the value of $e(v,\pi/2)$ is very strongly constrained.  In fact, its time derivative vanishes, as can be seen directly from the equations (\ref{eq:conserv1}).

In the microcanonical ensemble, the preferred black holes are the ones with the most entropy.  
In the $AdS_5 \times S^5$ case, ref.\ \cite{Dias:2016eto} demonstrated numerically that the spot-like solutions have higher entropy.  
Unfortunately, we cannot distinguish our black holes using entropy at leading order in a large $n$ expansion.  
At a fixed energy, all of our solutions have the same entropy because the two quantities are one and the same at leading order in $1/n$.
The entropy $S$ can be derived from the horizon area of the black hole using the Bekenstein-Hawking formula.  
In the large $n$ expansion, the leading term is given by
\begin{equation}
S \propto \int_0^\pi R_+ \sin^n \theta \, d \theta \ .
\end{equation}
The horizon location is the energy density in our notation, $R_+ = e(v, \theta)$.  
A similar feature was noticed in a large $d$ limit of planar black holes in ref.\ \cite{Herzog:2016hob}.
To calculate differences in entropy, we presumably need to calculate higher order terms in our large $n$ expansion.

The time evolution itself can indicate which configurations are preferred.  
In ref.\ \cite{Herzog:2016hob}, entropy production was encoded in second order derivative terms in the evolution equations. The type of allowed evolution was then constrained by the sign of the coefficients of these second order terms.
In our case, 
if generic initial conditions evolve to spot-like rather than belt-like black hole solutions, 
we could tentatively conclude that the spot-like solutions are preferred.   
(In practice, of course, we find no stable endpoint to the time evolution of the equations (\ref{eq:conserv1}) when $r_c^2 < 1/2$.)
A possible drawback is that the evolution equations (\ref{eq:conserv1}) are purely first order, and thus may not encode any type of entropy production.
The rescaled evolution equations (\ref{eq:conserv2}), on the other hand, have some second derivative terms.  

The rescaled equations (\ref{eq:conserv2}) also lead to energy conservation, i.e.\ that 
\begin{equation}
E_{\rm tot} = \int_{-\infty}^\infty e(v,u) e^{-u^2/2} du \ 
\end{equation}
is time independent,
applying the boundary conditions that $j = 0$ and $\partial_u e = 0$ as $|u| \to \infty$. 
At leading order in $n$, the entropy is again proportional to the total energy.  
Thus all the black holes connected by the evolution equations (\ref{eq:conserv2}) have the same entropy at leading order in $n$.
On the other hand, as these equations have second derivative terms, it may be more reliable to draw conclusions about the preferred
endpoint of the instability from the 
time evolution of generic initial data.

\section{Summary and outlook}
\label{sec:summary}

We have applied the large $d$ approximation to the study of black hole instability and its evolution in an $AdS_d \times S^d$ background. We derived analytic expressions for the instability threshold and quasinormal spectra order by order in the large $d$ expansion. The results agree with the numerical calculations. We also studied a nonlinear ansatz for the metric, restricted to fluctuate along three of the $2d$ coordinates: time, the polar angle of  the $S^d$, and the radial coordinate of the $AdS_d$. Simple hydrodynamic-like equations were found, one in the polar angle $\theta$ and the other zooming in on the equator of the sphere with $u = \left( \theta - \frac{\pi}{2} \right)\sqrt{n}$. 
We find that the systems admit solutions very like the lumpy black spots and belts numerically constructed in ref.\ \cite{Dias:2015pda}. 
We provided simple analytic formula for these solutions at leading order in the large $d$ limit. 
While the solutions in ref.\ \cite{Dias:2015pda,Dias:2016eto} were static, our solutions are time dependent.
In the stable regime $r_+^2 > 1/2$, fluctuations quickly relax to the homogeneous Schwarzschild solution.  
In the unstable regime $r_+^2 < 1/2$, we find exponentially growing solutions in the $\theta$ variable that do not approach a static endpoint.  
Despite their strong time dependence, the solutions do resemble at least qualitatively the static belt and spot solutions found in ref.\ \cite{Dias:2015pda,Dias:2016eto}.  In the rescaled $u$ coordinate, we can find a quasi-static solution.  At intermediate times, with $r_+^2 < 1/3$ below the threshold for the $\ell=2$ instability,  the energy density is well fit by the static belt-like solution that we found.  However, at long times, numerical evolution suggests this belt-like solution is unstable as well.

We would like to be able to figure out the ultimate fate of these unstable small black holes in $AdS_d \times S^d$.  The answer to this question has implications both for general relativity and field theory.  On the geometric side, it is interesting to understand what types of black hole horizon topologies are possible and how they may or may not arise through time evolution.  The behavior at large $d$ may shed light on
issues concerning cosmic censorship, in particular on when and whether black holes can break apart.  On the field theory side, through the AdS/CFT correspondence in the special case $d=5$, the behavior of these small black holes maps out the phase diagram of maximally supersymmetric four dimensional $SU(N)$ Yang-Mills theory at large $N$ and strong 't Hooft coupling in the micro-canonical ensemble.

\begin{figure}
\begin{center}
            \includegraphics[width=4in]{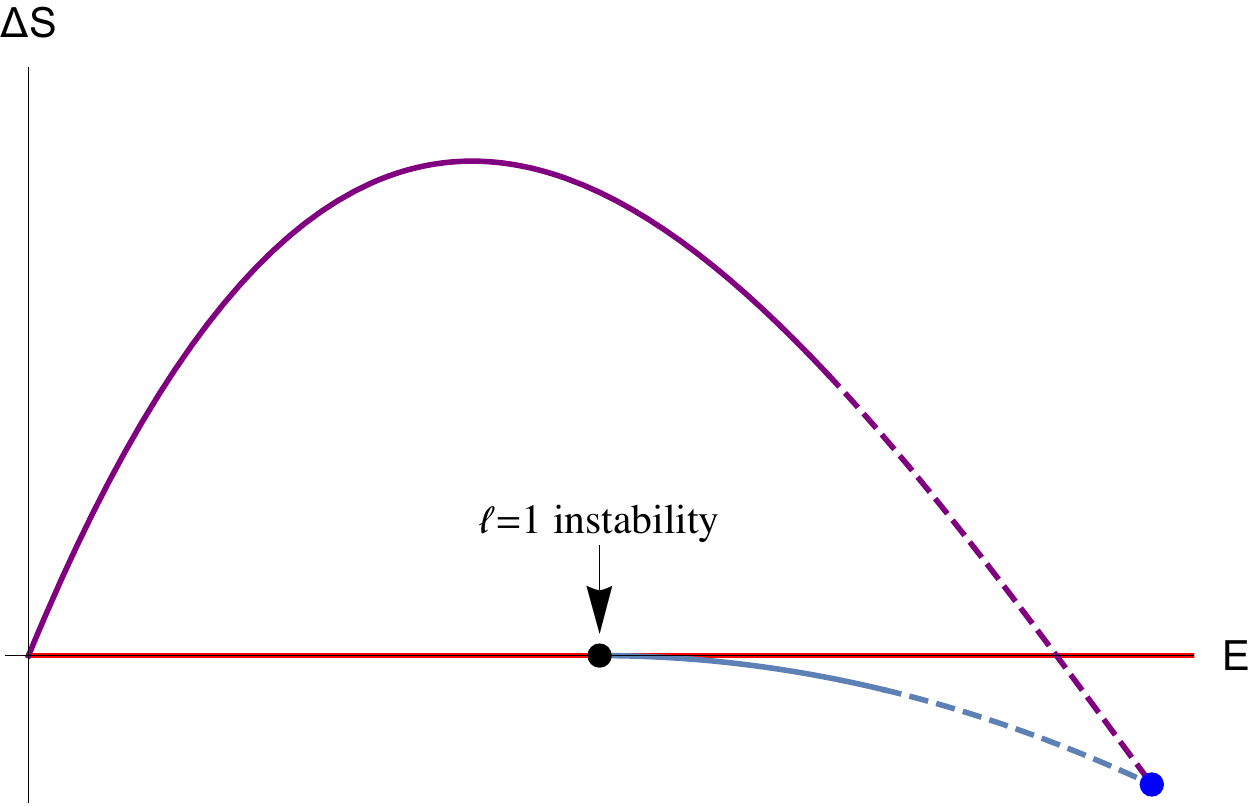}
\end{center}
 \caption{
 Change in entropy vs.\ energy plot for small black hole solutions in $d=5$, based on fig.\ 3 of ref.\ \cite{Dias:2016eto}.  The horizontal red line is the homogeneous Schwarzschild solution.  The black dot is the onset of the $\ell=1$ instability.  The solid blue line emanating from the $\ell=1$ instability indicates numerically constructed spot-like solutions with $S^3 \times S^5$ topology.  The solid purple line indicates numerically constructed spot-like solutions with $S^8$ topology.  The dashed lines are extrapolations.  
  }
\label{fig:microplot}
\end{figure}

Based on an analysis of static solutions in the $d=5$ case, refs.\ \cite{Dias:2015pda, Dias:2016eto} conclude that the branch of lumpy black holes  emanating from the $\ell=1$ instability at $r_+ = 0.4402$ have lower entropy and higher energy than the homogeneous Schwarzschild solution, and are thus disfavored in the micro-canonical ensemble.  However, they find an additional set of black hole solutions where the horizon has $S^8$ topology.  This set has greater entropy than the homogeneous solution, and an energy range that extends both above and below the $\ell=1$ instability.  
Based on an extrapolation of this family, they conjecture that there exists a first order phase transition to these $S^8$ black holes from the homogeneous solutions at an $r_+ > 0.4402$.  In fact, one can plausibly connect these $S^8$ solutions and lumpy black holes at a cusp, forming a classic swallowtail shape for a first order phase transition.  See figure \ref{fig:microplot}.

It is important to note that refs.\ \cite{Dias:2015pda, Dias:2016eto} construct static solutions with no time dependence.  While their conjectured phase transition is economical and reasonable, ideally one should be able to get to these static solutions through time evolution.  Without  evidence from time evolution, there remain other possibilities for the behavior of these small black holes \cite{Yaffe:2017axl}.   There could be other static or stationary solutions that preserve less symmetry on the $S^5$ or involve other fields from type IIB supergravity.  It could be that the instability never reaches any kind of static or even stationary final state, and the solution remains chaotically time dependent, or evolves to the string scale where general relativity ceases to be a useful description.  

With regard to the arguments in the previous two paragraphs, our analysis comes up short.  
We only determined the time evolution at leading order in $d$.  At this leading order, entropy and energy are the same.  At fixed energy, we cannot determine which of our homogeneous, spot-like, and belt-like solutions are preferred with respect to entropy, as they all have the same entropy.  We anticipate, however, that with some extra work subleading $1/d$ corrections to the entropy could be worked out and used to distinguish the different solutions.

We did not see dynamical evidence that stable spot-like solutions were preferred.  Our failure to see these spot-like solutions, however, should not be construed as evidence that a large $d$ generalization of the conjecture of refs.\ \cite{Dias:2015pda, Dias:2016eto}
is incorrect.  One possibility -- the most likely in our view -- is that the run-away behavior we did see could be stabilized by including $1/d$ corrections to the evolution equations.  As several run-away simulations involve the energy density growing uncontrollably at one of the poles of the $S^d$, it is plausible that stabilizing this run-away behavior will lead to a black hole with effectively $S^{2d-2}$ horizon topology at the corresponding pole. 
Another possibility is that the $1/d$ corrections will not stabilize the run-away behavior, and that general relativity will cease to be a useful description once the curvatures get sufficiently large.  
One thing we can tentatively rule out is fractal behavior at large $d$, as observed for the black string in \cite{Lehner:2010pn}.
We did not see any evidence for fractal time evolution in our large $d$ limit.  The energy density evolved reliably to form one or two lumps on the $S^d$ with little or no substructure.

Like most of the literature on this subject of small black holes in anti-de Sitter, our ansatz limits functional dependence to $\theta$, time, and the radius of $AdS_d$.  Moreover, we only turn on the metric and a self-dual $d$-form field strength.  We can say very little at this point 
about solutions which involve more complicated coordinate dependence.  We can also say very little about solutions which generalize to large $d$ type IIB supergravity solutions with more fields turned on. 
It should be straightforward and probably worthwhile 
to include an extra coordinate or two on the $S^d$ and $AdS_d$ and search for solutions that preserve less symmetry.  It could be true, for instance, that a path to a static solution at late times, even if that solution preserves an $SO(d)$ symmetry on the $S^d$, may at intermediate times require the presence of less than $SO(d)$ symmetry.  
 Generalizing type IIB supergravity to large $d$, on the other hand, seems like a challenging and less well-defined enterprise.

From the preceding discussion, working out $1/d$ corrections to our results is a key next step.
In addition to resolving some of the puzzles about entropy production and run-away behavior in the time evolution, one may hope that with enough
subleading terms in the expansion, the results may provide not just a qualitative understanding of the most interesting  
case $d=5$, but a quantitative one as well. 
Another important project for the future is to work out in more detail the entries of the holographic dictionary in this large $d$ limit.  For instance, how does the quantity $e(v, \theta)$ we called energy density relate to the actual energy density of the field theory through the usual prescription for calculating the stress tensor?  In the case $d=5$, the expectation values of scalar operators in the Yang-Mills theory can be read from the holographic stress-tensor \cite{Dias:2015pda, Dias:2016eto}.  Can something similar be done in this large $d$ limit?

\acknowledgments
We would like to thank Alex Buchel, Veronika Hubeny, Saebyeok Jeong, Luis Lehner, Shiraz Minwalla, Mukund Rangamani, Martin Ro\v{c}ek,  Michael Spillane, Benson Way, and Laurence Yaffe for useful discussions.
This work was supported in part by the National Science Foundation 
under Grant No.\ PHY-1620628.

\appendix

\allowdisplaybreaks

\section{Equation of motion for $\chi_2$}
Given the linearized ansatz (\ref{eq:conserv2}), the fluctuation of the $rr$ component of the metric, $\chi_2$, follows the differential equation
\begin{equation}
\chi_2'' + P \chi_2' + Q \chi_2 = 0.
\end{equation}
$P$ and $Q$ depend on $d, r,\omega^2, \ell, f$, and derivatives of $f$, where $f = 1 + r^2 - (r_+/r)^{d-3} (1+r_+^2)$. This equation in an arbitrary number of dimensions was inferred from solving Einstein's equations for several different specific values of $d$. All the other components follow from this equation; hence we may call it the master equation. $P$ and $Q$ are given by 
\begin{align*}
\hspace{-10mm}P =& \Bigg[
-(6-d) r^2 \myf^2 \mydf^2 \left(r \left(d (\ell+2)+\ell^2-\ell-2\right)-(d-2) \mydf\right)\\\nn
&+r^3 \myf \Big(-5(d-2) \omega^2 \mydf^2-2 (d-2) \mydf^4+2 r \mydf^3 \left(d (\ell+2)-\myddf+\ell^2-\ell-2\right) \\\nn
&\qquad  -4 (d-2) \omega^4+6 r \omega^2 \mydf \myddf \Big)\\\nn
&-r^2 \myf^2 \left(2 \mydf \left(r^2 \left(d (\ell+2)+\ell^2-\ell-2\right) \myddf+(d-2)^2
   \omega^2+r^2 \left(-\myddf^2\right)\right)-6 (d-2) r \omega^2 \myddf\right)\\\nn
&+4 r \myf^3 \Big( r^2 \left(d(\ell+2)+\ell^2-\ell-2\right) \myddf+(d-2) (d-1) \mydf^2\\\nn
&\qquad + r \mydf \left(\myddf-(d-1) \left(d
   (\ell+2)+\ell^2-\ell-2\right)\right)+(d-2) d \omega^2-r^2 \myddf^2 \Big)\\\nn
&+4 d \myf^4 \left(r \left(d(\ell+2)-\myddf+\ell^2-\ell-2\right)-(d-2) \mydf\right)\\\nn
&+(d+2) r^3 \myf^2 \mydf^2 \myddf-3 r^4 \omega^2 \mydf \left(\mydf^2+4 \omega^2\right) 
\Bigg] / \\\nn
&\Bigg[ r \myf \Bigg( 
r^2 \myf \left(-2 (d-2) \omega^2 \mydf-(d-2) \mydf^3+r \mydf^2 \left(d (\ell+2)-\myddf+\ell^2-\ell-2\right)+2 r \omega^2 \myddf\right)\\\nn
&\qquad-4 r \myf^2 \left(-(d-2) \mydf^2+r \mydf\left(d (\ell+2)-\myddf+\ell^2-\ell-2\right)-(d-2) \omega^2\right)\\\nn
&\qquad+4 \myf^3 \left(r \left(d(\ell+2)-\myddf+\ell^2-\ell-2\right)-(d-2) \mydf\right) -r^3 \omega^2 \left(\mydf^2+4 \omega^2\right)
\Bigg) \Bigg]
\end{align*}
and
\begin{align*}
\hspace{-10mm}Q =& \Bigg[
r^3 \omega^2 \myf \Big( 2 r \omega^2 \left(2 \left(d (\ell+2)+\ell^2-\ell-2\right)-3 \myddf\right)-14 (d-2) \omega^2 \mydf-5
   (d-2) \mydf^3\\\nn
& \qquad + r \mydf^2 \left(4 \left(d (\ell+2)+\ell^2-\ell-2\right)-3 \myddf\right) \Big)\\\nn
&+r^2 \myf^2 \Bigg( 2\omega^2 \left(r^2 \left(d (\ell+2)+\ell^2-\ell-2\right) \myddf+6 (d-2) \omega^2\right)-2 (d-2) d \mydf^4\\\nn
&\qquad - \mydf^2 \left(-r^2 \left(d (\ell+2)+\ell^2-\ell-2\right) \myddf+r^2 \left(d (\ell+2)+\ell^2-\ell-2\right)^2+4 (d-3) (d-2)
   \omega^2\right)\\\nn
&\qquad-2 r \omega^2 \mydf \left(-2 (d+2) \myddf-(d-10) \left(d (\ell+2)+\ell^2-\ell-2\right)\right)\\\nn
&\qquad+r \mydf^3 \left((3 d-2) \left(d (\ell+2)+\ell^2-\ell-2\right)-2 d \myddf\right) \Bigg)\\\nn
&+2 r \myf^3 \Bigg(r \omega^2 \left(2((d-3) d-1) \myddf-2 (d-5) \left(d (\ell+2)+\ell^2-\ell-2\right)\right)+3 (d-2) d \mydf^3\\\nn
&\qquad +\mydf \Big(-(d+2) r^2
   \left(d (\ell+2)+\ell^2-\ell-2\right) \myddf+d r^2 \myddf^2+2 r^2 \left(d (\ell+2)+\ell^2-\ell-2\right)^2\\\nn
&\qquad\qquad+2 (d-2)^2 \omega^2\Big)\\\nn
&\qquad -r \mydf^2 \left((5 d-4) \left(d (\ell+2)+\ell^2-\ell-2\right)-d (d+1) \myddf\right)\Bigg)\\\nn
&-4 \myf^4 \Big(r^2 \left(-(d+1) \left(d (\ell+2)+\ell^2-\ell-2\right) \myddf+d \myddf^2+\left(d
   (\ell+2)+\ell^2-\ell-2\right)^2\right)\\\nn
&\qquad +(d-2) d \mydf^2-(d-1) r \mydf \left(2 \left(d (\ell+2)+\ell^2-\ell-2\right)-d
   \myddf\right)\Big)\\\nn
&-r^4 \omega^2 \left(5 \omega^2 \mydf^2+\mydf^4+4 \omega^4\right)
\Bigg] / \\\nn
&\Bigg[ r \myf^2 \Bigg(
r^2\myf \left(-2 (d-2) \omega^2 \mydf-(d-2) \mydf^3+r \mydf^2 \left(d (\ell+2)-\myddf+\ell^2-\ell-2\right)+2 r \omega^2
   \myddf\right)\\\nn
&\qquad -4 r \myf^2 \left(-(d-2) \mydf^2+r \mydf \left(d (\ell+2)-\myddf+\ell^2-\ell-2\right)-(d-2)
   \omega^2\right)\\\nn
&\qquad + 4 \myf^3 \left(r \left(d (\ell+2)-\myddf+\ell^2-\ell-2\right)-(d-2) \mydf\right) - r^3
   \omega^2 \left(\mydf^2+4 \omega^2\right)
\Bigg)\Bigg] \ ,
\end{align*}
where we have defined $\mydf \equiv \partial_r f(r)$ and $\myddf \equiv \partial_r^2 f(r)$.

\label{app:chi2}

\section{Ricci tensor for a metric with two internal spaces}
\label{app:ricci}

In this appendix, we consider the Ricci tensor for the following type of metric
\begin{align}
ds^2 &= g_{\Pi\Sigma} dx^\Pi dx^\Sigma\\\nn
&= \bar{g}_{\alpha\beta} dx^\alpha dx^\beta + \exp(2A) \tilde{g}_{\theta_1\theta_2} dx^{\theta_1}dx^{\theta_2} + \exp(2B) \tilde{g}_{\phi_1\phi_2} dx^{\phi_1}dx^{\phi_2}
\end{align}
with the coordinate dependence
\begin{align}
\bar{g}_{\alpha\beta} &= \bar{g}_{\alpha\beta}(x^\gamma), \quad A = A(x^\gamma), \quad B = B(x^\gamma),\\\nn
\tilde{g}_{\theta_1\theta_2} &= \tilde{g}_{\theta_1\theta_2}(x^{\theta}), \quad \tilde{g}_{\phi_1\phi_2} = \tilde{g}_{\phi_1\phi_2}(x^{\phi}).
\end{align}
The metric $\bar{g}_{\alpha\beta}$ represents the base space, whose coordinates are indexed with $\alpha, \beta, \gamma, \cdots$. The metrics $\tilde{g}_{\theta_1\theta_2}$ and $\tilde{g}_{\phi_1\phi_2}$ represent the internal spaces fibered over the base space. Their coordinate indices are $\theta_1, \theta_2, \cdots$ and $\phi_1, \phi_2, \cdots$ with the numbers of dimensions $n_A$ and $n_B$. 
Capital letters run over the coordinates of the full space.

We will derive the expression for the Ricci tensor using the tetrad formalism, in which the computation is simpler than in the coordinate basis. In terms of the vielbeins, we can write the line element as
\begin{align}
ds^2 &= \eta_{MN} e^M \otimes e^N\\\nn
& = \eta_{ab} \bar{e}^a \otimes \bar{e}^b  + \exp(2A) \delta_{k_1 k_2} \tilde{e}^{k_1} \otimes \tilde{e}^{k_2}  + \exp(2B) \delta_{l_1 l_2} \tilde{e}^{l_1} \otimes \tilde{e}^{l_2} .
\end{align}
We use the Greek letters for the coordinate indices and the Latin letters for the vielbein indices. The vielbeins $\bar{e}^a, \tilde{e}^k,$ and $\tilde{e}^l$ are those of the base and internal spaces, whose elements satisfy
\begin{align}
\bar{g}_{\alpha\beta} = \tensor{\bar{e}}{_\alpha^a} \tensor{\bar{e}}{_\beta^b} \eta_{ab}, \quad
\tilde{g}_{\theta_1 \theta_2} = \tensor{\tilde{e}}{_{\theta_1}^{k_1}} \tensor{\tilde{e}}{_{\theta_2}^{k_2}} \delta_{k_1 k_2}, \quad
\tilde{g}_{\phi_1 \phi_2} = \tensor{\tilde{e}}{_{\phi_1}^{l_1}} \tensor{\tilde{e}}{_{\phi_2}^{l_2}} \delta_{l_1 l_2}.
\end{align}
In terms of these vielbeins for the base and the internal spaces, we may write the full vielbeins as (with the one-form bases explicit)
\begin{align}
\tensor{e}{_\Pi^a} dx^\Pi = \tensor{\bar{e}}{_\alpha^a} dx^\alpha, \quad
\tensor{e}{_\Pi^{k_1}} dx^\Pi = \exp(A) \tensor{\tilde{e}}{_{\theta_1}^{k_1}} dx^{\theta_1}, \quad
\tensor{e}{_\Pi^{l_1}} dx^\Pi = \exp(B) \tensor{\tilde{e}}{_{\phi_1}^{l_1}} dx^{\phi_1}.
\end{align}
Recall that, in the tetrad formalism, the torsion and the curvature two-form are defined as
\begin{align}
T^M &= de^M + \tensor{\omega}{^M_N} \wedge e^{N},\\\nn
\tensor{{\mathcal R}}{^M_N} &= d\tensor{\omega}{^M_N} + \tensor{\omega}{^M_P} \wedge \tensor{\omega}{^P_N}.
\label{eq:structure}
\end{align}
The torsion free condition $T^M = 0$, along with a bit of guessing, allows us to determine the spin connection $\tensor{\omega}{^M_N}$. Because of the symmetry between two internal spaces and the antisymmetry of two indices $M$ and $N$, we only need to determine
\begin{align}
\tensor{\omega}{^a_b}, \quad \tensor{\omega}{^a_{k_1}}, \quad \tensor{\omega}{^{k_1}_{k_2}}, \quad \tensor{\omega}{^{k_1}_{l_1}}.
\end{align}
These can be inferred from the $a$ and $k$ components of the torsion free condition. Writing out the $a$ component, we have
\begin{align}
de^a + \tensor{\omega}{^a_b} \wedge e^b + \tensor{\omega}{^a_{k_1}} \wedge e^{k_1} + \tensor{\omega}{^a_{l_1}} \wedge e^{l_1} = 0.
\end{align}
This equation is satisfied if we choose
\begin{align}
\tensor{\omega}{_\Pi^a_b} dx^\Pi = \tensor{\overline{\omega}}{_\alpha^a_b} dx^\alpha, \quad
\tensor{\omega}{^a_{k_1}} \propto e_{k_1},\quad
\tensor{\omega}{^a_{l_1}} \propto e_{l_1},
\label{eq:guess1}
\end{align}
where $\tensor{\overline{\omega}}{^a_b}$ is the spin connection on the base space, which satisfies the torsion free condition of its own. We also examine the $k$ component, which is
\begin{align}
de^{k_1} + \tensor{\omega}{^{k_1}_a} \wedge e^a + \tensor{\omega}{^{k_1}_{k_2}} \wedge e^{k_2} + \tensor{\omega}{^{k_1}_{l_1}} \wedge e^{l_1} = 0.
\end{align}
Again, we follow the obvious choice,
\begin{align}
\tensor{\omega}{_\Pi^{k_1}_{k_2}} dx^\Pi = \tensor{\tilde{\omega}}{_{\theta_1}^{k_1}_{k_2}} dx^{\theta_1},
\end{align}
where $\tensor{\tilde{\omega}}{^{k_1}_{k_2}}$ is the spin connection on the one of internal spaces. We are left with
\begin{align}
\exp(A) (\partial_\alpha A) \tensor{\tilde{e}}{_{\theta_1}^{k_1}} dx^{\alpha} \wedge dx^{\theta_1} 
+ \tensor{\omega}{^{k_1}_a} \wedge e^a
+ \tensor{\omega}{^{k_1}_{l_1}} \wedge e^{l_1} = 0.
\end{align}
Since $\tensor{\omega}{^{k_1}_a}$ is proportional to $e^{k_1} = \tensor{e}{_{\theta_1}^{k_1}} dx^{\theta_1}$ (\ref{eq:guess1}), we are forced to choose
\begin{align}
\tensor{\omega}{^{k_1}_a} = (\partial_\alpha A) \tensor{e}{^\alpha_a} e^{k_1}, \quad \tensor{\omega}{^{k_1}_{l_1}} = 0.
\end{align}
The rest of spin connections can be determined by symmetry. Collecting the non-vanishing spin connections, we have
\begin{align}
\tensor{\omega}{^a_b} = \tensor{\overline{\omega}}{^a_b}, 
\quad \tensor{\omega}{^{k_1}_{k_2}} = \tensor{\tilde{\omega}}{^{k_1}_{k_2}}, \quad
\tensor{\omega}{^{l_1}_{l_2}} = \tensor{\tilde{\omega}}{^{l_1}_{l_2}},\\\nn
\tensor{\omega}{^{k_1}_a} = \exp(A) (\partial_\alpha A) \tensor{\bar{e}}{^\alpha_a} \tilde{e}^{k_1}, \quad
\tensor{\omega}{^{l_1}_a} = \exp(B) (\partial_\alpha B) \tensor{\bar{e}}{^\alpha_a} \tilde{e}^{l_1}.
\end{align}
From this result, we can proceed with the second of the defining relations (\ref{eq:structure}) to obtain the curvature two-form. There are only four components that need to be determined, which are
\begin{align}
\tensor{{\mathcal R}}{^a_b}, \quad \tensor{{\mathcal R}}{^a_{k_1}}, \quad \tensor{{\mathcal R}}{^{k_1}_{k_2}}, \quad \tensor{{\mathcal R}}{^{k_1}_{l_1}}.
\end{align}
It is straightforward to show that
\begin{align}
\tensor{{\mathcal R}}{^a_b} &= \tensor{\overline{{\mathcal R}}}{^a_b},\\\nn
\tensor{{\mathcal R}}{^{k_1}_{k_2}} &= \tensor{\widetilde{{\mathcal R}}}{^{k_1}_{k_2}} - \exp(2A) (\partial_\alpha A) (\partial^\alpha A)  \tensor{\tilde{e}}{_{\theta_1}^{k_1}} \tilde{e}{_{\theta_2 k_2}} dx^{\theta_1} \wedge dx^{\theta_2},\\\nn
\tensor{{\mathcal R}}{^{k_1}_{l_1}} &= - \exp(A+B) (\partial_\alpha A) (\partial^\alpha B) \tensor{\tilde{e}}{_{\theta_1}^{k_1}} \tilde{e}{_{\phi_1 l_1}} dx^{\theta_1} \wedge dx^{\phi_1} \ ,
\end{align}
where $\tensor{\overline{{\mathcal R}}}{^a_b}$ and $\tensor{\widetilde{{\mathcal R}}}{^{k_1}_{k_2}}$ are the curvature two-forms of the base and internal spaces.
Writing out the $a k_1$ component reveals
\begin{align}
\tensor{{\mathcal R}}{^a_{k_1}} &= \partial_\Pi (\tensor{\omega}{_{\theta_1}^a_{k_1}}) dx^\Pi \wedge dx^{\theta_1} + \tensor{\omega}{_\alpha^a_b} \tensor{\omega}{_{\theta_1}^b_{k_1}} dx^\alpha \wedge dx^{\theta_1} + \tensor{\omega}{_{\theta_2}^a_{k_2}} \tensor{\omega}{_{\theta_1}^{k_2}_{k_1}} dx^{\theta_2} \wedge dx^{\theta_1}\\\nn
&= -[ \partial_\beta( (\partial_\alpha A) e^{\alpha a} \exp(A)) - \tensor{\omega}{^a_b} ( (\partial_\alpha A) \bar{e}^{\alpha b} \exp(A)) ] \tilde{e}_{\theta_1 k_1} dx^\beta \wedge dx^{\theta_1}\\\nn
&\hspace{3cm}-[ (\partial_\alpha A) e^{\alpha a} \exp(A) ][ \partial_{\theta_2} \tilde{e}_{\theta_1 k_1} + \tensor{\tilde{\omega}}{_{\theta_1}^{k_2}_{k_1}} \tilde{e}_{\theta_2 k_2}] dx^{\theta_2} \wedge dx^{\theta_1}\\\nn
&= - ( \overline{\nabla}_\alpha \partial_\beta A + (\partial_\alpha A)( \partial_\beta A)) \exp(A) e^{\alpha a} \tilde{e}_{\theta_1 k_1} dx^\beta \wedge dx^{\theta_1}
\end{align}
where, in the last step, we used the torsion free condition and the tetrad postulate,
\begin{align}
\partial_\mu \tensor{e}{_\nu^a} - \overline{\Gamma}^\lambda_{\mu\nu} \tensor{e}{_\lambda^a} + \tensor{\omega}{_\mu^a_b} \tensor{e}{_\nu^b} = 0.
\end{align}
As before, the bars indicate objects in the base space. From these results, we can compute the Ricci tensor in a coordinate basis by converting it from the vielbein basis,
\begin{align}
R_{\Pi \Sigma} = \tensor{R}{^\Omega_{\Pi \Omega \Sigma}} = \tensor{e}{^\Omega_M}\tensor{e}{_\Pi^N} \tensor{R}{^M_{N\Omega \Sigma}}.
\end{align}
With a bit of care in reading off the components of differential forms, and noticing that $(dx^{\theta_1} \wedge dx^{\theta_2})_{\theta_3 \theta_4} = \delta^{\theta_1}_{\theta_3}\delta^{\theta_2}_{\theta_4} - \delta^{\theta_1}_{\theta_4}\delta^{\theta_2}_{\theta_3}$, we finally find the Ricci tensor
\begin{align}
R_{\alpha\beta} &= \overline{R}_{\alpha\beta} - n_A ( \overline{\nabla}_\alpha \partial_\beta A + (\partial_\alpha A)( \partial_\beta A)) - n_B ( \overline{\nabla}_\alpha \partial_\beta B + (\partial_\alpha B)( \partial_\beta B)),\\\nn
R_{\theta_1 \theta_2} &= \widetilde{R}_{\theta_1 \theta_2} - \exp(2A) \tilde{g}_{\theta_1 \theta_2} ( \overline{\nabla}^\alpha \partial_\alpha A + n_A (\partial_\alpha A)( \partial^\alpha A) + n_B (\partial_\alpha A)( \partial^\alpha B)).
\end{align}
The components $R_{\phi_1 \phi_2}$ are the same as $R_{\theta_1 \theta_2}$ with simple exchanges, $\theta \leftrightarrow \phi$ and $A \leftrightarrow B$. Perhaps the most interesting part is the cross term $(\partial_\alpha A)( \partial^\alpha B)$. This term would have been absent if there had been only one internal space in the fibration, and therefore hard to guess a priori.

%
%

For the metric ansatz of this paper (\ref{eq:metric}), the Ricci tensor and scalar can be written as

\begin{align}
\label{eq:riccifull}
R_{\alpha\beta} &= \overline{R}_{\alpha\beta}
-\sum_{J \in (A,B)} n_J \left( \frac{\overline{\nabla}_\beta \partial_\alpha g_J}{2g_J} - \frac{(\partial_\alpha g_J )(\partial_\beta g_J)}{4g_J^2} \right)
,\\\nn
R_{ij} &= (n_A - 1) \tilde{g}_{ij} - g_A \tilde{g}_{ij} \left( \frac{\overline{\nabla}^2 g_A}{2g_A}  + (n_A-2) \frac{(\partial^\alpha g_A )(\partial_\alpha g_A)}{4g_A^2} + n_B \frac{(\partial_\alpha g_A)( \partial^\alpha g_B)}{4g_A g_B} \right),\\\nn
R_{mn} &= (n_B - 1) \tilde{g}_{mn} - g_B \tilde{g}_{mn} \left( \frac{\overline{\nabla}^2 g_B}{2g_B}  + (n_B-2) \frac{(\partial^\alpha g_B)( \partial_\alpha g_B)}{4g_B^2} + n_A \frac{(\partial_\alpha g_A)( \partial^\alpha g_B)}{4g_A g_B} \right),\\\nn
R &= \overline{R} 
+ \sum_{J \in (A,B)} \left( 
n_J(n_J-1)g_J^{-1}  - n_J \left( \frac{\overline{\nabla}^2 g_J}{g_J} + (n_J-3) \frac{(\partial^\alpha g_J )(\partial_\alpha g_J)}{4 g_J^2} \right)
\right) \\ 
\nn
%
& \quad 
- n_A n_B \frac{(\partial^\alpha g_A )(\partial_\alpha g_B)}{2g_A g_B},
\end{align}
where $\alpha, \beta$ denote $v, r,$ or $\theta$; $i,j$ denote coordinates on $S^{d-2}$; and $m, n$ denote coordinates on $S^{d-1}$.

\section{Large $n$ expansion of the metric and field strength}
\label{app:metric}

In this appendix, we present all the terms of the metric and $d$-form field that we have determined for the ansatz (\ref{eq:metric}) in the large $n$ expansion. It is convenient to introduce the one form $f_\alpha$ and two form $f_{\alpha\beta}$ on the base space parametrized by $(v, r, \theta)$ to describe the self-dual $F_d$ form. The self-duality of $F_d$ implies
\begin{align}
f_{vr} &= g_{\theta\theta}^{-1/2} (f_\theta - f_r g_{v\theta}),\\\nn
f_{v\theta} &= g_{\theta\theta}^{-1/2} (f_\theta g_{v\theta} - f_r g_{v\theta}^2 + f_r g_{\theta\theta} g_{vv} - f_v g_{\theta\theta}),\\\nn
f_{r\theta} &= g_{\theta\theta}^{-1/2} f_r,
\end{align}
Then the eight equations of motion for the metric can be neatly expressed as 
\begin{align}
\label{eq:seom1}
&2 R_{\alpha\beta} = f_\alpha f_\beta + f_{\alpha\gamma} \tensor{f}{_\beta^\gamma},\\\nn
&4 R_{ij} = g_{ij} f_{\alpha\beta} f^{\alpha\beta},
\quad 2 R_{mn} = g_{mn} f_\gamma f^\gamma,
\end{align}
where $\alpha,\beta$ run through $(v,r,\theta)$, and $i,j$ and $m,n$ are the indices on the $S^{d-2}$ and $S^{d-1}$. For the $d$-form field, we have the four constraints
\begin{align}
\label{eq:seom2}
0 &= \partial_\gamma ( g_A^{(d-2)/2} f_{\alpha\beta} ) dx^\gamma \wedge dx^\alpha \wedge dx^\beta,\\\nn
0 &= \partial_\alpha (g_B^{(d-1)/2} f_\beta) dx^\beta \wedge dx^\alpha.
\end{align}
These equations of motion are solved by the following metric and field solutions
%
\begin{align}
\label{eq:metric-form-sol}
g_{vv} &= -\left( 1 + r_c^2 R^{2/n} - \frac{e(v,\theta)}{R} \left( 1 + r_c^2 e(v,\theta)^{2/n} \right) \right) + \frac{g_{vv}^{(1)}}{n} + \frac{g_{vv}^{(2)}}{n^2} + \mathcal{O}(n^{-3}),\\\nn
g_{v\theta} &= \frac{j(v,\theta)}{nR} + \frac{g_{v\theta}^{(2)}}{n^2} + \mathcal{O}(n^{-3}),
\quad g_{\theta\theta} = 1 + \frac{1}{n^2} \frac{j^2}{R(1+r_c^2)e} + \mathcal{O}(n^{-3}), \\\nn
g_A &= r_c^2 R^{2/n}+ \frac{g_A^{(3)}}{n^3} + \mathcal{O}(n^{-4}), 
\quad g_B = \sin^2\theta + \frac{g_B^{(3)}}{n^3} + \mathcal{O}(n^{-4}),\\\nn
f_v &= \mathcal{O}(n^{-5/2}), \quad f_r = \frac{f_r^{(3/2)}}{n^{3/2}} + \mathcal{O}(n^{-5/2}),\\\nn
f_\theta &= \left( 2(n+2) \right)^{1/2} + \mathcal{O}(n^{-3/2}),\\\nn
\end{align}
where
\begin{align*}
g_{vv}^{(1)} &= -\frac{\log R}{R} j r_c \cot\theta   + \frac{e_1(v,\theta)}{R} \\\nn
g_{vv}^{(2)} &= \frac{\log R }{2R} \cot \theta \left[
-2r_c 
  (3  j +  j_1 + \tan \theta \, \partial_\theta j )
- r_c \log R \Big( j - \frac{r_c \cot \theta}{1+r_c^2} \frac{j^2}{e} \Big) \right] \\
& \quad \quad
+ \frac{e_2(v,\theta)}{R} + \frac{j^2}{2R^2} \\\nn
g_{v\theta}^{(2)} &= -\frac{r_c \cot \theta}{(1+r_c^2)} \frac{j^2}{e} \frac{\log R}{R}+ \frac{j_1(v,\theta)}{R} \\\nn
g_{B}^{(3)} &= - \sin^2 \theta \left( \frac{1}{r_c^2}g_A^{(3)} + \frac{1 }{R(1+r_c^2)} \frac{j^2}{e}  \right)\\\nn
f_{r}^{(3/2)} &= \frac{\sqrt{2} (\log R - \log e) j }{ (1+r_c^2)^2 (R-e) e } \left[ (1+r_c^2) e - r_c (\cot\theta) j \right]. \\\nn
\end{align*}
The $e(v,\theta)$ and $j(v,\theta)$ satisfy a pair of differential equations (\ref{eq:conserv1}), which appear as the $vv$ and $v\theta$ components of the Einstein equations at $n^1$ and $n^0$ orders. In addition, the $vv$ component at $n^0$ order also gives a differential equation for $e_1(v,\theta)$ and $j_1(v,\theta)$. These solutions solve all of the metric and $d$-form equations of motion up to $n^0$ order.

\vskip 0.1in
\noindent
{\bf Near the equator:} We also provide the field and metric solutions in the rescaled polar angle $u$, which is related to $\theta$ as\footnote{%
We can use an alternative rescaling, $e^{-u^2/2} = \sin^n \theta$, which is at leading order $\theta = \pi/2 + u/\sqrt{n} + \mathcal{O}(n^{-3/2})$ and leads to the same effective equations for $e$ and $j$.
}
\begin{equation}
\theta = \pi/2 + u/\sqrt{n}.
\end{equation}
We perform the coordinate transformation $R = (r/r_c)^n$ explicitly in the ansatz. Our ansatz is then
\begin{align}
ds^2 &= g_{vv}dv^2 + 2 r_c n^{-1} R^{1/n-1} dvdR + 2g_{vu}dvdu + g_{uu}du^2 + g_A d\Omega_{d-2}^2 + g_B d\Omega_{d-1}^2,\\\nn
F_d &= \left(g_A^{(d-2)}\det(S^{d-2})\right)^{1/2} \left(f_{\alpha\beta}/2 \right) dx^\alpha \wedge dx^\beta \wedge d\Omega_{d-2} + \left(g_B^{(d-1)} \det(S^{d-1})\right)^{1/2} f_\alpha dx^\alpha \wedge d\Omega_{d-1},
\end{align}
where $\alpha,\beta$ run over $(v, R, u)$. The effective equations for $e$ and $j$ are the same whether we express the ansatz in the $r$ or $R$. However, some of the Einstein equations are pushed to higher orders in $1/n$ in this setup, leaving fewer equations to solve. The self-duality of $F_d$ implies
\begin{align}
f_{vR} &= g_{uu}^{-1/2} ( - f_R g_{vu} + n^{-1} r_c R^{1/n-1} f_u ),\\\nn
f_{vu} &= g_{uu}^{-1/2} \left( n r_c^{-1} R^{1-1/n} f_R(-g_{vu}^2 + g_{vv}g_{uu}) + f_u g_{vu} - f_v g_{uu} \right), \\\nn
f_{Ru} &= g_{uu}^{1/2} f_R.
\end{align}
Then the equation of motions (\ref{eq:seom1}) and (\ref{eq:seom2}) are solved by the following metric and field expansion
\begin{align}
g_{vv} &= -(1 + r_c^2 R^{2/n}) \left( 1 - \frac{e(v,u)}{R} \right) + \frac{g_{vv}^{(1)}}{n} + \mathcal{O}(n^{-2}),\\\nn
g_{vu} &= \frac{j(v,u)}{nR} + \frac{g_{vu}^{(2)}}{n^2} + \mathcal{O}(n^{-3}), \quad g_{uu} = \frac{1}{n} + \frac{1}{n^2} \frac{j^2}{R(1+r_c^2)e} + \mathcal{O}(n^{-3}), \\\nn g_A &= r_c^2 R^{2/n} + \frac{g_A^{(2)}}{n^2} + \mathcal{O}(n^{-3}), \quad g_B = \cos^2(u/n^{1/2}) + \frac{g_B^{(2)}}{n^2} + \mathcal{O}(n^{-3}),\\\nn
f_v &= \mathcal{O}(n^{-2}), \quad f_r = \mathcal{O}(n^{-2}),\\\nn
f_u &= \left( \frac{2(n+2)}{n} \right)^{1/2} - \frac{g_B^{(2)}}{\sqrt{2}n} + \mathcal{O}(n^{-3}).\\\nn
\end{align}
where
\begin{align*}
g_{vv}^{(1)} &= \frac{j^2}{2R^2} + \frac{r_c \log R}{R}(-2r_c e + u j - \partial_u j) - \frac{e_1(v,u)}{R},\\\nn
g_{vu}^{(2)} &= \frac{j^3}{2R^2(1+r_c^2)e} 
- \frac{r_c \log R}{R(1+r_c^2)} \exp\left(\frac{u^2}{2}\right) \partial_u \left( \frac{j^2}{e} \exp\left(-\frac{u^2}{2}\right) \right)
- \frac{j_1(v,u)}{R},\\\nn
g_B^{(2)} &= -\frac{g_A^{(2)}}{r_c^2} - \frac{j^2}{R(1+r_c^2)e}.
\end{align*}
The $e(v,u)$ and $j(v,u)$ obey the effective equations of motion (\ref{eq:conserv2}), which are obtained from the $vv$ and $vu$ components of the Einstein equations at the $n^1$ and $n^0$ order. These solutions solve most of the equations of motion up to $n^0$ order. The equations that are not solved in terms of $(e,j,e_1,j_1)$ are the $vv$, $ij$, and $mn$ components of the Einstein equations (\ref{eq:seom1}) at $n^0$ order. The $ij$ and $mn$ components are the same, 
\begin{equation}
\partial_R \left[ R(R-e) \partial_R g_A^{(2)} \right] - \frac{2r_c^2}{1+r_c^2} g_A^{(2)} - \frac{r_c^2}{R^2(1+r_c^2)} j^2 = 0,
\end{equation}
which can be solved with standard techniques.
%

We are then left with the $vv$ component of the Einstein equations at $n^0$ order, which consists of $g_{vv}^{(2)}, g_{uu}^{(3)}, g_A^{(2)}, g_A^{(3)}, g_B^{(3)}$, and their first and second derivatives with respect to $R$, along with $e, j, e_1$, and $j_1$. The expression is too long to solve explicitly, but given the freedom to choose the aforementioned functions, we believe that it can be solved consistently with the boundary conditions.

\section{Spherical harmonics in the large $d$ limit}
\label{app:spher}

We show that there exist a class of spherical harmonics that depend only on the polar angle $\theta$ of $S^d$ and that tend to 
$Y_\ell \sim (\cos\theta)^\ell$ in the large $d$ limit. In general, the spherical harmonics on the sphere $S^d$ are given by
\begin{equation}
Y_{\ell_1 ... \ell_d} (\theta_1, \ldots, \theta_d) = \frac{1}{\sqrt{2 \pi}} e^{i \ell_1 \theta_1} \prod_{j=2}^{d} {}_j\overline{P}^{\ell_{j-1}}_{\ell_j} (\theta_j)
\end{equation}
where
\begin{equation}
{}_j\overline{P}^{m}_{\ell} (\theta) = \sqrt{\frac{2\ell+j-1}{2}\frac{(\ell+m+j-2)!}{(l-m)!}} (\sin\theta)^{-(j-2)/2} P^{-\ell-(j-2)/2}_{\ell+(j-2)/2} \left( \cos\theta \right).
\end{equation}
$P^m_\ell$ is the associated Legendre polynomial.  The spherical harmonics are eigenfunctions of the Laplacian: $\Delta Y_{\ell_1 ... \ell_d} =-\ell_d (\ell_d+d-1) Y_{\ell_1 ... \ell_d}$. Our nonlinear metric ansatz assumes a dependence only on $\theta=\theta_d$. We can find such $Y_{{\ell_1 ... \ell_d}}$ by setting $\ell_1, \ldots, \ell_{d-1}$ to zero and $\ell = \ell_d$. Then
\begin{equation}
Y_{\ell} (\theta) \propto {}_d\overline{P}^0_\ell (\theta) \propto \sin^{-(d-2)/2} P^{-(d-2)/2}_{\ell+(d-2)/2} \left( \cos\theta \right).
\end{equation}
But by the Rodrigues' formula (denoting $m=\frac{d-2}{2}$),
\begin{equation}
P^{-m}_{\ell+m} (x) \propto (1-x^2)^{m/2} \frac{d^{\ell+2m}}{dx^{\ell+2m}} (x^2-1)^{\ell + 2m}.
\end{equation}
As the number of dimensions becomes large, $d \rightarrow \infty$, the coefficient of $x^\#$ with the highest power dominates. As a result, in the large $d$ limit, the spherical harmonics in this class reduce to
\begin{equation}
Y_{\ell}(\theta) \propto (\cos\theta)^\ell + O(d^{-1}).
\end{equation}



\end{document}